\documentclass[12pt]{article}

\begin{document}

\begin{center}
{\large\bf NEW APPROACH TO DESCRIPTION OF MAJORANA PROPERTIES
	OF NEUTRAL PARTICLES}

\smallskip

{\bf Yu.V. Gaponov}\footnote{e-mail: gaponov2@imp.kiae.ru}

\smallskip

RRC "Kurchatov Institute",  Kurchatov sqr., 1,

Moscow, 123182 Russia

\end{center}

\begin{abstract}

\par Two mathematical models based on Pauli transformations
including U(1) chiral group and Pauli SU(2) group, that mixes
particle and antiparticle states, are developed for description of
Majorana properties of neutral particles. The first one describes
a system, incorporating left- and right-handed fermions of the
same flavor, and it is a generalization of the Majorana model
of his pioneer article of 1937 year. The second describes
a two-flavor neutrino system with quantum numbers of
Zel'dovich - Konopinsky - Mahmoud (ZKM) type. For massless
fermions the Pauli symmetry is exact and leads to the conserved
generalized lepton charge. It is a Pauli isospace vector, whose
different directions are coordinated with Dirac or generalized
Majorana properties. In nonzero - mass case the models describe
the combined Dirac - Majorana properties of neutral particles,
which are characterized either by the generalized lepton charges of
ZKM - type or by the eigenvalues of the operator that is the product
of the charge operator and chirality. The latter is connected with
operator of the structure of Lagrangian mass term or with the generalized
flavor number of the second model. The choice of the basic operator
depends on the inversion classes (A-B or C-D - types) of the particles
with respect to the space inversion. The modified second model can be
used for description of neutrino oscillation in the simplest two -
flavor case.

\end{abstract}

\par Investigation of Majorana properties of neutral particles \cite{1}
with neutrino as the most principal of them all is the chief goal of
modern weak processes physics. An examination of the properties was
essentially stimulated by the recent discovery of oscillations in
atmospheric, solar and reactor neutrino fluxes \cite{2} - \cite{4}. Majorana
models for neutrino description were developed in literature in two basic
versions: either in schemes proceeded from Majorana and Pontecorvo
\cite{1, 5} (in simplest case including one particle with left-handed and
right-handed states) or in phenomenological models with two and more
left-handed neutrinos of different species (flavors) \cite{6, 7} (see
also e.g. \cite{8} - \cite{13}) which are used today for analysis of
the neutrino oscillations . However the absence of quantum characteristics
suitable for the description of Majorana - type states is an essential
disadvantage of the models.

\par Meanwhile a new approach to construction of Majorana schemes
where the problem is partly solved is possible as it was demonstrated
by the author in \cite{14} - \cite{17}. It is based on
application of general Pauli (chiral - Pauli) transformations group
\cite{18} (which was Pauli - Pursey group in the terms of 1950's
years \cite{19} - \cite{21}). The use of the group allows to
construct some examples of phenomenological models in which quantum
numbers for description of Majorana states can be introduced.
Below two simplest models of such type which are in line
with above indicated versions will be presented: the model for one
neutral particle with left-handed and right-handed states of the same
flavor and a two-flavor neutrino model that includes states of different
flavors, for example, electron and muon neutrino. The latter will be
used then for description of neutrino oscillations.

\par The plan of the article is as follows: in $\S 1$ the Pauli
group terms are introduced, in $\S 2$ and $\S 3$ the above
indicated models are presented and $\S 4$ is consequently dedicated
an application of the modified two-flavor model to describe the
neutrino oscillations. Finally, in $\S 5$ we shall discuss peculiarities
of Pauli models which make them different from Majorana neutrino
phenomenological schemes presented in the literature.

\par {\bf $\S 1$. Introduction}

\par As it was demonstrated for the first time by Pauli \cite{18}
the fermion fields of zero mass are symmetric relatively to the following
transformations:
\begin{equation}
\begin{array}{c}
\psi^{'}(x) = e^{i\gamma_{5} \chi/2} (a \psi(x) +
b \gamma_{5} \gamma_{2} \gamma_{4} \overline{\psi}^{T}(x))
= e^{i\gamma_{5} \chi/2} (a \psi(x) +
b \gamma_{5} \psi^{C}(x)), \\
|a|^2 + |b|^2 = 1, \ \ \psi^{C}(x) = \hat{C} \overline{\psi}^{T}(x) =
\eta_{C} \gamma_{2} \gamma_{4} \overline{\psi}^{T}(x), \ \
(\eta_{C} = 1 \ \ \cite{22}).
\end{array}
\label{m1}
\end{equation}
\noindent These transformations involve pure Pauli SU(2) - group
(type I of Pauli \cite{18}) and chiral U(1) - transformations
(type II) and conserve commutation relations of the fields. The
former one for $a = e^{i\varphi/2}, \  b=0 \ $ can be reduced to the
phase transformation subgroup on whose base the conserved lepton
charge can be put into the scheme. With introduction of $\hat{\kappa}_{i}
 \ (i=x, y, z)$ operators and the generalized two-component function
$\Psi(x)$ one receives:
\begin{equation}
\begin{array}{c}
\hat{\kappa}_{x} = {0 \ +1 \choose  +1 \ \ 0}, \
\hat{\kappa}_{y} = {0 \ -i \choose  +i \ \ 0}, \
\hat{\kappa}_{z} = {+1 \ \ 0 \choose  0 \ -1}, \\

\Psi(x) = {\psi(x) \choose  \gamma_{5} \psi^{C}(x)}.
\end{array}
\label{m2}
\end{equation}
\noindent It is readily shown that for $a = e^{i\varphi/2} \cos{\theta/2},
 \ b = e^{i\varphi/2}e^{-i\phi}\sin{\theta/2} \ $ transformation (\ref{m1})
can be reduced to the standard form:
\begin{equation}
\begin{array}{c}
\Psi^{'}(x) =
e^{i\gamma_{5} \chi/2} e^{i\hat{\kappa}_{z} \varphi/2}
e^{i(\cos{\phi} \hat{\kappa}_{y} -\sin{\phi}\hat{\kappa}_{x})
\theta/2}\Psi(x) = S(\chi) S(\varphi) S(\phi, \ \theta)\Psi(x), \\

\hat{\kappa} = \cos{\theta}\hat{\kappa}_{z} +
\sin{\theta}\cos{\phi} \hat{\kappa}_{x}
+ \sin{\theta} \sin{\phi} \hat{\kappa}_{y}, \ \
S^{+}(\phi, \ \theta)\hat{\kappa}_{z} S(\phi, \ \theta)
= \hat{\kappa},
\end{array}
\label{m3}
\end{equation}
\noindent which can be interpreted in terms of rotations in  chiral
and Pauli subspaces. The latter includes $S(\varphi) =
e^{i\hat{\kappa}_{z}\varphi/2}$ rotations about $\kappa_{z}$ axis and
$S(\phi, \ \theta) = S$ transformation of $\vec{\kappa}_{z}$ vector
to the direction of $\vec{\kappa}$, defined by the standard Euler
angles $\phi$, $\theta$. Some other representations of the generalized
function are also possible. In this case the (\ref{m3}) form of Pauli
transformations should be consequently modified.

\par The analysis of the conservation condition of Pauli transformations
under $CPT$ - operation \cite{22, 23} shows (see details in \cite{16})
that $CPT$ - invariance leads to the following relation between
phases $\eta_{P}$, $\eta_{T}$ of discrete P- and T- transformations,
introduced in coordination with the standard definition (see \cite{22}),
 and $b$ parameter of Pauli transformations:
\begin{equation}
\begin{array}{c}
b(1+ \eta_{P}^{2} / \eta_{T}^{2}) = 0, \ \ or \
\ \eta_{P}^{2}/\eta_{T}^{2} = -1 \ \ for \ b \ \neq 0.
\end{array}
\label{m4}
\end{equation}
\noindent The latter relation is in accordance with the well known
condition $\eta_{P}^{2} = -1, \\ \eta_{T}^{2} = +1 $ adopted usually for
physical particles \cite{22}. Fermions of $\eta_{P} = \pm i$ will be
called  particles of inversion of A - and B - classes \cite{24}. However
 for Majorana particles one can not a priori exclude the other possible
choice $\eta_{P} = \pm 1$, that Racah, for the first time, called attention
to (see \cite{25} and also \cite{26, 27}). We shall describe such fermions
as particles of inversion of C - and D - classes \footnote{There is no
prescribed designation of inversion classes of the particles in literature,
here we are based on definitions of monograph \cite{24}, which are however
opposite to designations of \cite{26}} and treat them hypothetical
particles for which $\eta_{P}^{2} = +1, \ \ \eta_{T}^{2} = -1 $.

\par The inversion classes of neutral particles are connected with
their Majorana properties. Indeed, let us consider the reasonably general
 conditions of Majorana type in the following forms:
\begin{equation}
\begin{array}{c}
\psi^{C}(x, \zeta) =\lambda e^{i\phi} \psi(x, \zeta) \ \ (A), \ \
\psi^{C}(x, \zeta) =\lambda e^{i\phi} \gamma_{5} \psi(x, \zeta) \ \ (B),

 \\ ( \lambda - \ arbitrary \ real \ number, \ \zeta \ - \ quantum \
numbers \ of \ states),
\end{array}
\label{m5}
\end{equation}
\noindent It is easily shown that the former (\ref{m5}) $(A)$ condition
is fullfilled for particles of inversion A-B - classes only while the
latter (\ref{m5}) $(B)$ is fullfilled for C-D - ones. Note that the
limitation of (\ref{m5}) $(A)$ condition is used in modern Majorana models
\cite{10} - \cite{13}. It means that belonging of the particles to inversion
A-B - classes is implicitly included in such models. However we shall not
put this limitation into operation in this work and let the both capabilities
be realized. In doing so the general Majorana conditions (\ref{m5}) $(B)$
should be used for particles of C-D - classes.

\par
\par
{\bf $\S 2.$ Majorana properties of a neutral free fermion}

\par The fundamental existence of Majorana properties of a neutral
fermion was firstly de\-mon\-strat\-ed by E.Maj\-o\-ra\-na. Bas\-ing
on Dirac equa\-tion he show\-ed that it has a special solution that meets
condition $\psi^{C}(x) = \psi(x)$ \cite{1}. From the modern viewpoint he
constructed a particular solution that is a superposition of particle
and antiparticle states for the simplest case of one particle. We shall
reveal below that basing on Pauli transformations one can generalize
this result and construct a Majorana model for particles of one type
that contains solutions obeying similar but more general conditions
(\ref{m5}) $(A), (B)$. Simultaneously a generalized concept of lepton
charge will appear in the model. We shall initially develop the model
for massless particles and then for the case of massive ones
(this model was firstly proposed and investigated by the author in
\cite{14} - \cite{16}).

\par Let us examine a fermion field of zero mass which is symmetric under
Pauli transormations (\ref{m1}). The Lagrangian of the field presented
in generalized functions (\ref{m2}) is as follows:
\begin{equation}
\begin{array}{c}
L_{0}(x) = -\frac{1}{2}
[ \overline{\Psi}(x) \gamma_{\mu} \partial_{\mu} \Psi(x) ], \ \
\Psi(x) = {\psi(x) \choose  \gamma_{5} \psi^{C}(x)},
\end{array}
\label{m6}
\end{equation}
\noindent (here, wave functions are of secondary quantization). Note that
the chosen form of generalized functions is universal and it is
invariant with respect to Pauli transormations (\ref{m1}). The invariance
of the Lagrangian under chiral $S(\chi)$ and phase $S(\varphi)$
transitions leads to the conserved chiral and lepton charges:
\begin{equation}
\begin{array}{c} Q^{CH} = \frac{1}{2}
\int d^{3}x [\Psi^{+}(x) \gamma_{5} \Psi(x) ] \\
= \frac{1}{2} \int d^{3}x [\psi^{+}(x) \gamma_{5} \psi(x) +
\psi^{C +}(x) \gamma_{5} \psi^{C}(x) ] , \\

Q^{L} = Q^{P}_{z} = \frac{1}{2}
\int d^{3}x [\Psi^{+}(x) \hat{\kappa}_{z} \Psi(x)] = \\
\frac{1}{2} \int d^{3}x [\psi^{+}(x) \psi(x) -
\psi^{C +}(x) \psi^{C}(x) ].
\end{array}
\label{m7}
\end{equation}
\noindent They stipulate some characteristics of basic eigenfunctions
$\Psi_{0}(x)$, those are chirality $\rho = \pm 1 \ (L, \ R)$ and
lepton charge $q_{z} = \pm 1$:
\begin{equation}
\begin{array}{c}
\gamma_{5} (\Psi_{0})_{\rho, q_{z}}(x) =
\rho (\Psi_{0})_{\rho, q_{z}}(x), \ \
\hat{\kappa}_{z} (\Psi_{0})_{\rho, q_{z}}(x) =
q_{z} (\Psi_{0})_{\rho, q_{z}}(x), \\

(\Psi_{0})_{\rho, q_{z} = +1}(x) =
{\psi_{0 \rho}(x) \choose 0}, \ \

(\Psi_{0})_{\rho, q_{z} = -1}(x) =
{0 \choose \gamma_{5} \psi^{C}_{0 \rho}(x)} =
{0 \choose \rho \psi^{C}_{0 \rho}(x)}.
\end{array}
\label{m8}
\end{equation}
\noindent As it follows from (\ref{m7}) in the space of Pauli
transformations the lepton charge operator $\hat{\kappa}_{z}$
is connected with a vector oriented along z - axis in a way that
particle solution $q_{z} = + 1$ and antiparticle one
$q_{z} = - 1$ are matched with two possible projections of
the vector to z - axis. The conservation of the lepton charge
is evident from invariance of the Lagrangian under rotations about
the vector. In Pauli scheme this vector describes the lepton charge.
Note that in the case of massless particle there are no physical
reasons for separation of z-axis in comparison with other directions.

\par Such a peculiarity of Pauli scheme lets one to introduce
a concept of generalized lepton charge $Q^{P}$ coordinated with
an arbitrarily chosen direction of Pauli isospace. In this case its z -
component conserves the connection with lepton charge $Q^{P}_{z}$
and x - , y - components occur to coordinate with Majorana
properties of the particle. The generalized lepton charge
can be obtained from $Q^{P}_{z}$ by using Pauli transormation
$S^{+}(\phi, \ \theta)$ of $\Psi_{0}(x)$ wave function and
$\hat{\kappa}_{z}$ operator. It takes the form:
\begin{equation}
\begin{array}{c}
Q^{P} = \frac{1}{2}\int d^{3}x \Psi^{+}(x)\hat{\kappa}\Psi(x)

= \cos{\theta}Q^{P}_{z} + \sin{\theta}(\cos{\phi}Q^{P}_{x} +
\sin{\phi}Q^{P}_{y}) = \\

\frac{1}{2} \int d^{3}x [\cos{\theta} (\psi^{+}(x) \psi(x) -
\psi^{C +}(x) \psi^{C}(x)) + \\ \sin{\theta}
(\psi^{+}(x) e^{-i\phi} \gamma_{5} \psi^{C}(x) +
\psi^{C +}(x) e^{+i\phi} \gamma_{5} \psi(x)) ],   \\

\hat{\kappa} = \cos{\theta} \hat{\kappa}_{z} + \sin{\theta}
(\cos{\phi} \hat{\kappa}_{x} + \sin{\phi} \hat{\kappa}_{y}), \ \

\Psi(x) = S^{+}(\phi, \ \theta) \Psi_{0}(x),
\end{array}
\label{m9}
\end{equation}
\noindent and is coordinated with $\kappa$ direction defined
in Pauli iso\-space by stan\-dard Euler angles $(\phi, \ \theta)$.
The eigenfunctions of the generalized lepton charge operator with
$\rho, \ q = \pm 1$ quantum numbers are constructed from the
generalized functions $\Psi_{0}(x)$ (\ref{m8}) by using the
same transformations:
\begin{equation}
\begin{array}{c}
\hat{\kappa} \Psi_{\rho, q}(x) = q \Psi_{\rho, q}(x), \\

\Psi_{\rho, +1}(x) =
{\psi_{\rho, +1}(x) \choose
\gamma_{5} \psi^{C}_{\rho, +1}(x) }
= {\cos{(\frac{\theta}{2})} \psi_{0 \rho}(x) \choose
e^{i\phi} \sin{(\frac{\theta}{2})} \psi_{0 \rho}(x) }, \\

\Psi_{\rho, -1}(x) =
{\psi_{\rho, -1}(x) \choose  \gamma_{5} \psi^{C}_{\rho, -1}(x)}
= {- \rho e^{-i\phi}\sin{(\frac{\theta}{2})} \psi^{C}_{0 \rho}(x)
\choose \rho \cos{(\frac{\theta}{2})} \psi^{C}_{0 \rho}(x)},
\end{array}
\label{m10}
\end{equation}
\noindent They describe consequently two independent solutions of
 different signs $q = \pm 1$. With fixed values of the generalized
lepton charge q there are following relations between components
 $\psi^{C}_{\rho, q}(x)$ and $\psi_{\rho, q}(x)$ of new system of
the eigenfunctions conserving their universal (\ref{m2}) form:
\begin{equation}
\begin{array}{c}
\psi_{q}^{C}(x) =
q(\tan{(\frac{\theta}{2})})^{q} e^{i\phi} \gamma_{5} \psi_{q}(x),
 \ (q = \pm 1), \ \

\psi_{q}(x) = \sum_{\rho} \psi_{\rho, q}(x), \\

\Psi_{q}(x) = \sum_{\rho} \Psi_{\rho, q}(x) =
{\psi_{q}(x) \choose
\gamma_{5} \psi^{C}_{q}(x)}, \ \
\int d^{3} x \Psi_{q}^{+}(x) \Psi_{q'}(x) = \delta_{q, q'},
\end{array}
\label{m11}
\end{equation}
\noindent  They are projection conditions which set up the
de\-fi\-ni\-te eigen\-valu\-es q and cor\-res\-pond\-ing
ei\-gen\-func\-tions of the
generalized lepton charge. These conditions have a form of
modified Majorana relations of (\ref{m5}) $(B)$ - type and
are fulfilled only in a case when a particle under
investigation belongs to inversion C-D - classes. It is seen that
the choice of generalized lepton charge operator as the basic
one is coordinated with definite inversion classes of the particle.

\par For description of particles of A-B - inversion classes
it is necessary to choose the product of chirality and
generalized lepton charge operator as basic operator. In doing so
it is useful to represent the generalized wave function in a new
universal form, that is also invariant under arbitrary chiral - Pauli
transformations. In the case the general form of these changes
of transformations (\ref{m3}) is following:
\begin{equation}
\begin{array}{c} \Phi(x) = {\psi_{L}(x) +
\eta \psi^{C}_{R}(x) \choose \psi_{R}(x) + \eta \psi^{C}_{L}(x)}, \\
\Phi^{'}(x) =
e^{i\gamma_{5} \chi/2} e^{i\hat{\kappa}_{z} \gamma_{5} \varphi/2}
e^{i\eta (\cos{\phi}\hat{\kappa}_{y} -
\sin{\phi}\hat{\kappa}_{x}\gamma_{5}) \theta/2} \Phi(x) = \\
S(\chi) S'(\varphi) S'(\phi, \theta) \Phi(x).
\end{array}
\label{m12}
\end{equation}
\noindent In the new representation, the Pauli isospace is modified.
Now it is based on $\kappa_{x}\gamma_{5}, \  \kappa_{y}, \
\kappa_{z}\gamma_{5}$ basic vectors coordinated with x, y, z - axes
that corresponds to the form of Pauli rotations and lepton charge
operators. In the basic case correspondent with (\ref{m8}) the
lepton charge operator takes $\hat{\kappa}_{z}\gamma_{5}$ form
and its product with the chirality operator is described by
$\hat{\kappa}_{z}$ form. In this representation the eigenfunctions
of $\kappa_{z} = \rho q_{z}$ quantum numbers and $\eta = \pm 1$
charge parity have the following form:
\begin{equation}
\begin{array}{c}
\hat{\kappa}_{z} (\Phi_{0})_{\kappa_{z}, \eta}(x) =
\kappa_{z} (\Phi_{0})_{\kappa_{z}, \eta}(x); \\

(\Phi_{0})_{\kappa_{z} = +1, \eta}(x) =
{\psi_{0 L}(x) + \eta \psi^{C}_{0 R}(x) \choose 0}, \ \
(\psi_{0 R}(x) = 0); \\

(\Phi_{0})_{\kappa_{z} = -1, \eta}(x) =
{0 \choose \psi_{0 R}(x) + \eta \psi^{C}_{0 L}(x)}, \ \

(\psi_{0 L}(x) = 0).
\end{array}
\label{m13}
\end{equation}
\noindent The lepton charge in $\Phi_{0}(x)$ state is defined by
the relation:
\begin{equation}
\begin{array}{c}
Q^{L} = Q^{P}_{z} =
\int d^{3}x

\Phi_{0}^{+}(x)\hat{\kappa}_{z} \gamma_{5} \Phi_{0}(x).
\end{array}
\label{m14}
\end{equation}
\noindent The functions (\ref{m13}) describe two Majorana solutions
of $\eta$ charge parity constructed either of the left particles and
their right antiparticles ($\kappa_{z} = + 1$) or of the right
particles and their left antiparticles ($\kappa_{z} = - 1$).
Inserting them in (\ref{m14}) one can readily make sure that the mean
value of the lepton charge is equal to zero in these states. It is
evident that $\hat{\kappa}_{z}$ operator coordinates with z - axis
of the modified Pauli space and two Majorana solutions (\ref{m13})
coordinate with two projections onto z - axis of a vector, created
as a product of chirality and the lepton charge vector. In the
general case, that can be derived from (\ref{m13}) by
using Pauli rotation $S'^{+}(\phi, \theta)$, the eigenfunctions of
$\hat{\kappa}$ operator, which correspond to the product of chirality
and the generalized lepton charge, are as follows:
\begin{equation}
\begin{array}{c}
\hat{\kappa} \Phi_{\kappa, \eta}(x) = \kappa \Phi_{\kappa,\eta}(x), \\

\Phi_{+1, \eta}(x) =
{\cos{(\frac{\theta}{2})}
(\psi_{0 L}(x) + \eta \psi^{C}_{0 R}(x))
\choose
\eta e^{i\gamma_{5} \phi} \sin{(\frac{\theta}{2})}
(\psi_{0 L}(x) + \eta \psi^{C}_{0 R}(x)) }; \\

\Phi_{-1, \eta}(x) =
{- \eta e^{-i\gamma_{5} \phi} \sin{(\frac{\theta}{2})}
(\psi_{0 R}(x) + \eta \psi^{C}_{0 L}(x))
\choose \cos{(\frac{\theta}{2})}
(\psi_{0 R}(x) + \eta \psi^{C}_{0 L}(x)) },
\end{array}
\label{m15}
\end{equation}
\noindent The generalized lepton charge is consequently as follows:
\begin{equation}
\begin{array}{c}
Q^{P}(\kappa) =

\int d^{3}x \Phi^{+}_{\kappa, \eta}(x) \hat{\kappa} \gamma_{5}

\Phi_{\kappa, \eta}(x),  \\

\hat{\kappa} \gamma_{5} = \cos{\theta}\hat{\kappa}_{z}\gamma_{5} +
\eta \sin{\theta}(\cos{\phi}\hat{\kappa}_{x}\gamma_{5} +
\sin{\phi}\hat{\kappa}_{y}).
\end{array}
\label{m16}
\end{equation}
\noindent These expressions describe a pair of Majorana solutions
$\kappa = \pm 1$ in terms of the previous system of eigenfunctions.
Obviously that for the states with fixed $\kappa$ quantum number
the mean value of the generalized charge tends to zero in the general
case as well. For fixed $\kappa$ values the following connections
arise between separate chiral components $\psi_{\rho, \kappa}(x)$
and $\psi^{C}_{\rho, \kappa}(x)$ entering into the universal form of
the generalized function (\ref{m12}):
\begin{equation}
\begin{array}{c}
\Phi_{\kappa, \eta}(x) =

{\psi_{L \kappa}(x) + \eta \psi^{C}_{R \kappa}(x)

\choose
\psi_{R \kappa}(x) + \eta \psi^{C}_{L \kappa}(x) }, \\

\psi_{\rho, \kappa }^{C}(x) =
\kappa(\tan{(\frac{\theta}{2})})^{\rho \kappa}

e^{i\phi} \psi_{\rho, \kappa}(x) \ \ (\rho = \pm 1 \ (L, \ R)),
\end{array}
\label{m17}
\end{equation}
\noindent They have the form of generalized Majorana conditions of
(\ref{m5}) $(A)$ type. They are projection conditions which set off
the definite eigenvalues of $\hat{\kappa}$ operator as it was in
case (\ref{m11}), however, here they are consistent with the case
when particles under investigation are of inversion A-B - classes.
Thus, it occurs that in general case of Pauli scheme the choice of
basic operators using for description of eigenfunctions of the
states occurs to be directly correlated with inversion classes of
the particles. In case of their being of inversion C-D - classes the
generalized lepton charge operator should be chosen as a basic
one. And if they are of inversion A-B - classes its product with
chirality operator should be chosen. The projection conditions
to set off fixed eigenfunctions of the basic operator, have the form
of general ones of either (\ref{m5}) $(B)$ or $(A)$ types. While
the considerations have been made on the base of investigation of
massless case the approach will also hold for particles of nonzero
mass with some exception as we shall show below.

\par Let us turn now to the description of massive particles and
state the general concept for this case. In accordance with
views of Standard Model the masses of particles are determined by
mean vacuum value of Higg's field that has its origin in spontaneous
breakdown of a basic symmetry proposed for them with their masses
being zero. The introduction of the mass terms in Lagrangian
(\ref{m6}) that is symmetric under Pauli transformations (\ref{m1})
is connected with the breakdown of the symmetry. It is possible to
suppose that in the model under investigation the Pauli symmetry
plays a role of the basic symmetry. Basing on example of a particular
Dirac case one can see that the breakdown leads to production
of separated directions in chiral and Pauli subspaces. If the mechanism
of the broken symmetry is universal, the Higg's vacuum values, because
of different subspace directions, are connected after breakdown
by the same transformations as the directions themselves.
Basing on Dirac case with a well known mechanism of mass generation,
one can construct in the same manner the Lagrangian mass
terms of the general form including both Dirac and Majorana
terms for the model under investigation. As a result the Pauli scheme
for massive free particles, which in basic features is in coincidence
with standard phenomenological scheme of the modern Majorana
models \cite{8} - \cite{13}, arises and it occurs to be a particular
case of the models. However, it has some important peculiarities
due to Pauli transformations which it is based on. The scheme is
presented just below (for more details see \cite{15} - \cite{16}).

\par Let us begin from the Dirac case with describing it in representation
(\ref{m12}) of generalized wave functions. In the representation
the Lagrangian of Dirac type, Dirac equation and corresponding lepton
charge have the following forms:
\begin{equation}
\begin{array}{c}
L(x) = L_{0}(x) -
\frac{M}{2} \overline{\Phi}_{D}(x)\hat{\kappa}_{x} \Phi_{D}(x) , \\
Q^{L} = Q^{P}_{z} =
\frac{1}{2}\int d^{3}x \Phi^{+}_{D}(x)\hat{\kappa}_{z}
\gamma_{5} \Phi_{D}(x), \\

(\gamma_{\mu} \partial_{\mu} + M \hat{\kappa}_{x})\Phi_{D}(x) = 0, \
\Phi_{D}(x) = {\psi_{DL}(x) + \eta \psi^{C}_{DR}(x) \choose
\psi_{DR}(x) + \eta \psi^{C}_{DL}(x)}.
\end{array}
\label{m18}
\end{equation}
\noindent As one can see from these expressions the invariance of Dirac
Lagrangian under chiral and pure Pauli groups of general Pauli
transformations is broken. The source of the breakdown is in the mass
term of Lagrangian. However, it does not break the invariance under
the phase subgroup of Pauli rotations about z-axis that is connected
with the conserved lepton charge $Q^{L} = Q^{P}_{z}$. The expression
of the lepton charge through generalized Dirac functions
$\Phi_{D}(x)$ is similar to (\ref{m14}) for massless fermions.
Consequently it does not depend on mechanism of the breakdown
introduced by mass terms. In the chosen representation the operator of
lepton charge is coordinated with z - axis of the Pauli isospace
while the operator of structure of the mass term is coordinated with
x - axis. Thus in representation (\ref{m12}) for generalized wave
functions in the Dirac case the spontaneous breakdown
introduced by the mass term, leads to the production of two different
isolated directions: z - axis of Pauli rotations responsible for
the existence of conserved lepton charge and x - direction, whose
operator determining the structure of the mass term is connected with.

\par It is easy to verify that the mass term of the Dirac Lagrangian
is not sensitive to the inversion class of a particle. For this
reason the alternative choice for the basic operator of
representation of eigenfunctions is possible in the Dirac case: itis
either the lepton charge operator or its product with chirality.
The latter one occurs to be connected with the operator of the structure
of the Lagrangian mass term.

\par The representation (\ref{m18}) is in compliance with the
choice of lepton charge operator. The eigenfunctions of
$q_{z} = \pm 1$ charge take the following forms in this case:
\begin{equation}
\begin{array}{c}
\hat{\kappa}_{z}\gamma_{5}\Phi_{D, q_{z}}(x) =
q_{z} \Phi_{D, q_{z}}(x), \\

\Phi_{D, q_{z} = +1}(x) =
{\psi_{D L}(x) \choose \psi_{D R}(x)}, \ \
\Phi_{D, q_{z} = -1}(x) =
{\eta \psi^{C}_{D R}(x) \choose \eta \psi^{C}_{D L}(x) }, \\
\int d^{3}x \Phi_{D, q_{z}}^{+}(x) \Phi_{D, q'_{z}(x)} =
\delta_{q_{z}, q'_{z}}.
\end{array}
\label{m19}
\end{equation}
\noindent Such a representation of solutions is equivalent to a normal
description of a particle as "Dirac" like ("Dirac neutrino"),
when the particle and antiparticle states differ by the lepton charge
values. An alternative representation can be deduced from (\ref{m18})
by Pauli rotation $S'^{+}(\phi = 0, \theta = - \eta \pi/2)$ of
wave function $\Phi_{D}(x)$. Then the Dirac Lagrangian, the Dirac
equation and the lepton charge are transformed to the form:
\begin{equation}
\begin{array}{c}
L(x) = L_{0}(x) -
\frac{M}{2} \overline{\Phi}_{MD}(x)\hat{\kappa}_{z} \Phi_{MD}(x) , \ \

(\gamma_{\mu} \partial_{\mu} + M\hat{\kappa}_{z})\Phi_{MD}(x) = 0, \\

Q^{P} = - \frac{1}{2}
\int d^{3}x \Phi^{+}_{MD}(x)\hat{\kappa}_{x}\gamma_{5} \Phi_{MD}(x), \\

\Phi_{MD}(x) = e^{i \hat{\kappa}_{y} \pi/4}\Phi_{D}(x) =

\frac{1}{\sqrt{2}} {\psi_{DL}(x) + \psi_{DR}(x)
+ \eta \psi^{C}_{DR}(x) + \eta \psi^{C}_{DL}(x))
\choose
\psi_{DR}(x) - \psi_{DL}(x) +
\eta \psi^{C}_{DL}(x) - \eta \psi^{C}_{DR}(x)) }.
\end{array}
\label{m20}
\end{equation}
\noindent The basic operator $\hat{\kappa}_{z}$ defining
the structure of the mass term in the representation is
in compliance with product ($\hat{\kappa}_{z}\gamma_{5}$)
(which is the basic operator of the charge representation (\ref{m18})),
and ($\gamma_{5}$) chirality. It is coordinated with z -axis and
its eigenfunctions, specified by $\kappa_{z}= \pm 1, \ \eta = \pm 1$
eigenvalues, are as follows:
\begin{equation}
\begin{array}{c}
\hat{\kappa}_{z}(\Phi_{MD})_{\kappa_{z}, \eta}(x) =
\kappa_{z}(\Phi_{MD})_{\kappa_{z}, \eta}(x);\\

(\Phi_{MD})_{+1, \eta}(x) =
\frac{1}{2}{\psi_{D}(x) + \eta \psi^{C}_{D}(x) \choose 0}  =
{\psi_{D L}(x) + \eta \psi^{C}_{D R}(x) \choose 0}  =
{\psi_{MD 1}(x) \choose 0}, \\
(\psi^{C}_{D L (R)}(x) = \eta \psi_{D L (R)}(x)); \\

(\Phi_{MD})_{-1, \eta}(x) =
\frac{1}{2}{0 \choose -\gamma_{5}(\psi_{D}(x) - \eta \psi^{C}_{D}(x))} =
{0 \choose \psi_{D R}(x) + \eta \psi^{C}_{D L}(x)} =
{0 \choose \psi_{MD 2}(x)}, \\
(\psi^{C}_{D L (R)}(x) = - \eta \psi_{D L (R)}(x));

\end{array}
\label{m21}
\end{equation}
\noindent and can be interpreted in terms of two independent Majorana
particles $\psi_{MD 1}(x),  \psi_{MD 2}(x)$. They generate
a complete set of solutions similar to a particle - antiparticle
pair, so that every solution can be represented as a superposition
of two latter ones and vice - versa. (Note that (\ref{m21})
fits the secondary quantized form and the normalization of
the eigenfunctions, as compared with (\ref{m20}), changes when
some additional Majorana conditions are taken into account.) The
eigenvalue $\kappa_{z} = +1$ describes the Majorana solution of
the $\eta$ charge parity, being set off $\psi^{C}_{D}(x) = \eta
\psi_{D}(x)$ projection condition, and $\kappa_{z} = -1$ one describes
the solution of the same charge parity, resulting from the condition
$\psi^{C}_{D}(x) = - \eta \psi_{D}(x)$. The mean value of the
lepton charge for each of them is
equal to zero. These solutions are known as "Majorana solutions"
("Majorana neutrino"). The first of them (for $\eta = +1$) was
suggested for the first time by Majorana in \cite{1}.

\par Therefore, there are two alternative solutions in the Dirac case
due to a different choice of the isolated directions in the Pauli isospace
with a spontaneous symmetry breakdown being responsible for the
appearence of the mass terms. In the charge ("Dirac") representation the
lepton charge is connected with a vector, directed along the z - axis,
and the mass term alignes with a vector along the x - axis. In "Majorana"
(mass) representation the situation is contrary. Here, the z - axis is
aligned with the vector connected with the operator of structure of
the Lagrangian mass term and the lepton charge is aligned with
the x - axis. The Pauli transformation connects them by the rotation
about the y - axis to the $\frac{\pi}{2}$ angle. Owing to the
independence of the Dirac Lagrangian from inversion classes, both
alternative representations can be used for a description of physical
particles of the inversion A-B - classes as well as hypothetical
particles of C-D - classes.

\par In contrast with the Dirac case the choice of the basic operator of
representation in the general case of the Pauli scheme is essentially
connected with inversion classes of the particle under investigation.
In conventional designations the most general form of the Lagrangian
for the Pauli model and its generalized charge are as follows:
\begin{equation}
\begin{array}{c}
L(x) =  L_{0}(x) - \frac{M}{2} \{\cos{\theta}(

e^{+i\chi} \overline{\psi}_{R}(x)\psi_{L}(x) +
e^{+i\chi} \overline{\psi}_{L}^{GC}(x)\psi^{GC}_{R}(x) + \\
e^{-i\chi} \overline{\psi}_{L}(x)\psi_{R}(x) +
e^{-i\chi} \overline{\psi}_{R}^{GC}(x)\psi^{GC}_{L}(x))

+ \sin{\theta}(
\overline{\psi}_{R}(x)\psi_{L}^{GC}(x)  + \\
\overline{\psi}^{GC}_{L}(x)\psi_{R}(x)  -
\overline{\psi}^{GC}_{R}(x)\psi_{L}(x)  -
\overline{\psi}_{L}(x)\psi_{R}^{GC}(x) ) \}, \\

Q^{P}  =  \frac{1}{2}\int d^{3}x [\cos{\theta}
(\psi^{+}_{L}(x) \psi_{L}(x) - \psi^{GC +}_{L}(x) \psi^{GC}_{L}(x)  +
\psi^{+}_{R}(x) \psi_{R}(x) - \\
\psi^{GC +}_{R}(x) \psi^{GC}_{R}(x))

+ \sin{\theta} (\psi^{+}_{L}(x) e^{-i\chi} \psi^{GC}_{L}(x) +
\psi^{GC +}_{L}(x) e^{+i \chi} \psi_{L}(x) - \\
\psi^{+}_{R}(x) e^{i\chi} \psi^{GC}_{R}(x) -
\psi^{GC +}_{R}(x) e^{-i\chi} \psi_{R}(x))],  \\

(\psi_{L}(x))^{GC} = \psi_{R}^{GC}(x) =
e^{-i(\chi + \phi)}\psi_{R}^{C}(x), \\
(\psi_{R}(x))^{GC} = \psi_{L}^{GC}(x) =
e^{+i(\chi - \phi)}\psi_{L}^{C}(x),
\end{array}
\label{m22}
\end{equation}
\noindent Here, the latter expressions define the designations
due to a new discrete operation of the generalized charge
conjugation (GC - conjugation). In fact, it is easy to verify
that the Lagrangian (\ref{m22}) is invariant under the following
transformation:
\begin{equation}
\begin{array}{c}
\psi_{L}(x) \leftrightarrow \psi^{GC}_{R}(x), \ \
\psi_{R}(x) \leftrightarrow \psi^{GC}_{L}(x),
\end{array}
\label{m23}
\end{equation}
\noindent which changes the sign of the generalized charge $Q^{P}$.
There are two phase factors in the definition of GC - conjugation
operation: the universal one depending on $\phi$ angle,
whose introduction is equivalent to entering the eneral phase factor
$\eta_{C} = e^{-i\phi}$ into the definition of the standard charge
conjugation operation (\ref{m1}), and an extra one which depends
on chiral characteristics of a particle and is a peculiarity of the
Pauli scheme. The latter takes into account a fact that the phase
factors $\eta_{C}$ for left- and right - handed particles can pricipally
be different. By comparing the general Pauli Lagrangian (\ref{m22})
with the existing Majorana models \cite{10} - \cite{12}, \cite{29}
it is easy to verify that the case in question is connected with
the breakdown of CP - conservation.

\par Note a peculiarity of mass terms of the general Lagrangian and
the generalized charge (\ref{m22}) that is typical of the model under
investigation. Their Dirac parts (proportinal to $\cos{\theta}$)
are scalar under the space reflection independently of inversion
classes of particles. However the similar properties of their
Majorana terms (proportinal to $\sin{\theta}$) are essentially
dependent on the inversion classes. For C-D - classes of particles
the Majorana terms are scalar, but for A-B - classes some of them
are scalar and the others are pseudoscalar. Consequently the choice
of a basic operator of the representation using for particles
eigenfunctions description turns to be dependent in general
on their inversion classes.

\par In move from the Dirac case (\ref{m18}) to the general form of
the charge representation using the generalized lepton charge as
a basic operator, the Lagrangian and the generalized charge take the
following forms:
\begin{equation}
\begin{array}{c}
L(x) = \\ L_{0}(x) - \frac{M}{2} \overline{\Phi}^{GC}(x)
[\cos{\theta} (\cos{\chi}\hat{\kappa}_{x} +
\sin{\chi}\hat{\kappa}_{y})
- \eta \sin{\theta}\hat{\kappa}_{z}] \Phi^{GC}(x),  \\

Q^{P} = \frac{1}{2} \int d^{3}x \Phi^{GC +}(x)
[\cos{\theta}\hat{\kappa}_{z} +  \eta \sin{\theta}
(\cos{\chi}\hat{\kappa}_{x} + \sin{\chi}\hat{\kappa}_{y})]
\gamma_{5} \Phi^{GC}(x).
\end{array}
\label{m24}
\end{equation}
\noindent GC - functions entered into expressions are consisted
with the form (\ref{m12}), with an operator of C - conjugation being
exchanged by GC - one and with having their own Pauli transformations
law of the following form:
\begin{equation}
\begin{array}{c}
\Phi^{GC}(x) =
{\psi_{L}(x) + \eta \psi^{GC}_{R}(x)

\choose \psi_{R}(x) + \eta \psi^{GC}_{L}(x)} =
{\psi_{L}(x) + \eta e^{- i(\chi + \phi)} \psi^{C}_{R}(x)
\choose \psi_{R}(x) + \eta e^{+ i(\chi - \phi)} \psi^{C}_{L}(x)}, \\

\Phi'^{GC}(x) =
e^{i\eta (\cos{\chi}\hat{\kappa}_{y} - \sin{\chi}\hat{\kappa}_{x})
\theta/2} e^{i\hat{\kappa}_{z} \gamma_{5} \varphi/2}
e^{i\gamma_{5} \chi/2} \Phi^{GC}(x)  = \\
S'(\chi, \theta) S'(\varphi) S(\chi) \Phi^{GC}(x).
\end{array}
\label{m25}
\end{equation}
\noindent It is essentially to underline that with the use of GC - functions
the values of (\ref{m24}) expressions can be newly interpreted as vectors
of the basic Pauli isospace constructed on $\kappa_{x}, \ \kappa_{y},
\ \kappa_{z}$ vectors related to x, y, z - axes. In this case, the
expressions (\ref{m24}) contain $\chi$ and $\theta$ angles only
while the $\phi$ angle, after its having been introduced into
the definition of the generalized GC - function, is not already
explicitly included.

\par The eigenfunctions of the fixed generalized lepton charge, being
originated from (\ref{m19}) as a result of the general Pauli transformation,
take the following forms:
\begin{equation}
\begin{array}{c}
[\cos{\theta}\hat{\kappa}_{z} +  \eta \sin{\theta}
(\cos{\chi}\hat{\kappa}_{x} + \sin{\chi}\hat{\kappa}_{y})]
\gamma_{5} \Phi_{q}(x) = q \Phi_{q}(x), \\

\Phi_{q} (x) =
S'^{+}(\chi, \theta)e^{-i\hat{\kappa}_{z}(\chi/2)}\Phi_{D, q_{z}}(x), \ \
(q = q_{z}), \\
\Phi^{GC}_{q = +1}(x) =
{\psi_{L}(x) + \eta \psi^{GC}_{R}(x)
\choose \psi_{R}(x) + \eta \psi^{GC}_{L}(x)}_{q= +1} = \\

{e^{-i\chi/2} \cos{(\theta/2)} \psi_{D L}(x) -
\eta e^{-i\chi/2} \sin{(\theta/2)} \psi_{D R}(x)
\choose
e^{+i\chi/2} \cos{(\theta/2)} \psi_{D R}(x) +
\eta e^{+i\chi/2} \sin{(\theta/2)} \psi_{D L}(x)}, \\

\Phi^{GC}_{q = -1}(x) =
{\psi_{L}(x) + \eta \psi^{GC}_{R}(x)

\choose \psi_{R}(x) + \eta \psi^{GC}_{L}(x)}_{q = -1} = \\

{- e^{-i\chi/2} \sin{(\theta/2)} \psi^{C}_{D L}(x) +
\eta e^{-i\chi/2} \cos{(\theta/2)} \psi^{C}_{D R}(x)
\choose
e^{+i\chi/2} \sin{(\theta/2)} \psi^{C}_{D R}(x) +
\eta e^{+i\chi/2} \cos{(\theta/2)} \psi^{C}_{D L}(x)}.
\end{array}
\label{m26}
\end{equation}
\noindent As it follows from these relations, the modified projection
Majorana conditions typical of the inversion C-D - classes particles
are fulfilled for separate components of these eigenfunctions with
the quantum numbers $q = \pm 1$:
\begin{equation}
\begin{array}{c}
\psi^{C}_{q}(x) = \sum_{\rho} \psi^{C}_{q, \rho}(x) =
q (\tan{(\theta/2)})^{q} e^{i\phi} \gamma_{5} \psi_{q}(x).
\end{array}
\label{m27}
\end{equation}

\par A different situation arises in a process of generalization of
the Majorana representation (\ref{m20}) when the structure operator of
the Lagrangian mass terms is chosen as basic. Then the general Lagrangian
and the charge in terms of GC - functions take the forms:
\begin{equation}
\begin{array}{c}
L(x) = L_{0}(x) - \frac{M}{2} \overline{\Phi}^{GC}_{M}(x)
[\cos{\theta}\hat{\kappa}_{z} +
\eta \sin{\theta}(\cos{\chi} \hat{\kappa}_{x} +
\sin{\chi} \hat{\kappa}_{y})] \Phi^{GC}_{M}(x),  \\

Q^{P} = - \frac{1}{2} \int d^{3}x \Phi^{GC +}_{M}(x)
[\cos{\theta} (\cos{\chi} \hat{\kappa}_{x} + \sin{\chi} \hat{\kappa}_{y})
- \eta \sin{\theta} \hat{\kappa}_{z}]
\gamma_{5} \Phi^{GC}_{M}(x),  \\

\Phi^{GC}_{M}(x) =
e^{i(\cos{\chi}\hat{\kappa}_{y}
- \sin{\chi}\hat{\kappa}_{x}) \pi/4} \Phi^{GC}(x) = \\

\frac{1}{\sqrt{2}} {\psi_{L}(x) + e^{-i\chi}\psi_{R}(x) +
\eta \psi^{GC}_{R}(x) + \eta e^{-i\chi} \psi^{GC}_{L}(x)
\choose
\psi_{R}(x) - e^{+i\chi}\psi_{L}(x) +
\eta \psi^{GC}_{L}(x) - \eta e^{i\chi} \psi^{GC}_{R}(x)}.
\end{array}
\label{m28}
\end{equation}
\noindent The correspondent eigenfunctions of mass structure term
operator have in general case the following form:
\begin{equation}
\begin{array}{c}
[\cos{\theta}\hat{\kappa}_{z} +
\eta \sin{\theta}(\cos{\chi} \hat{\kappa}_{x} +
\sin{\chi} \hat{\kappa}_{y})] \Phi^{GC}_{M \kappa}(x)
= \kappa \Phi^{GC}_{M \kappa}(x),  \\

\Phi^{GC}_{M \kappa = +1}(x) =

\frac{1}{\sqrt{2}}{\psi_{L}(x) + e^{-i\chi}\psi_{R}(x) +
\eta \psi^{GC}_{R}(x) + \eta e^{-i\chi} \psi^{GC}_{L}(x)
\choose
\psi_{R}(x) - e^{+i\chi}\psi_{L}(x) + \eta \psi^{GC}_{L}(x) -
\eta e^{i\chi} \psi^{GC}_{R}(x)}_{\kappa = +1} = \\

{e^{-i\chi/2} \cos{(\theta/2)} (\psi_{DL}(x) + \psi_{DR}(x))
\choose
e^{+i\chi/2} \eta \sin{(\theta/2)} (\psi_{DL}(x) + \psi_{DR}(x))},  \\

\Phi^{GC}_{M \kappa = -1}(x) =

\frac{1}{\sqrt{2}}{\psi_{L}(x) + e^{-i\chi}\psi_{R}(x) +
\eta \psi^{GC}_{R}(x) + \eta e^{-i\chi} \psi^{GC}_{L}(x)
\choose
\psi_{R}(x) - e^{+i\chi}\psi_{L}(x) + \eta \psi^{GC}_{L}(x) -
\eta e^{i\chi} \psi^{GC}_{R}(x)}_{\kappa = -1} = \\

{e^{-i \chi/2} \eta \sin{(\theta/2)} (\psi_{DL}(x) - \psi_{DR}(x))
\choose
e^{+i\chi/2} \cos{(\theta/2)} (\psi_{DR}(x) - \psi_{DL}(x))}.
\end{array}
\label{m29}
\end{equation}
\noindent In the case of $\Phi^{GC}_{M \kappa}(x)$ expressions,
the formulae meet projection conditions leading to the
Majorana conditions given below. The latter are typical of
the inversion A-B - classes particles and do not depend on the
chiral angle:
\begin{equation}
\begin{array}{c}
\psi^{C}_{\kappa, L}(x) = \kappa \eta
(\tan{(\eta \frac{\theta}{2} + \frac{\pi}{4})})^{\kappa}
e^{i\phi}\psi_{\kappa, L}(x),  \\

\psi^{C}_{\kappa, R}(x) = \kappa \eta
(\cot{(\eta \frac{\theta}{2} + \frac{\pi}{4})})^{\kappa}
e^{i\phi}\psi_{\kappa, R}(x), \\

\psi^{C}_{\kappa, \rho}(x) = \kappa \eta
(\tan{(\eta \frac{\theta}{2} + \frac{\pi}{4})})^{\kappa \rho}
e^{i\phi}\psi_{\kappa, \rho}(x), \ \ \rho = \pm 1 (L, \ R).
\end{array}
\label{m30}
\end{equation}
\noindent Thus, the representations (\ref{m24}) and (\ref{m28}) are
two possible general forms of description of the particles in
the model under investigation. The choice of the basic operator of
the representation (\ref{m24}) or (\ref{m28}) forms is defined by
the inversion class of the particles. For A-B - classes
(physical) particles it is an operator that specifies the structure of
the Lagrangian mass term. On the contrary for the C-D - classes
(hypothetical ones) particles it is an operator of the generalized
lepton charge. It is easy to verify that the former (described in
its basic representation) can be presented as a product of the
latter one (in its own representation) and the operator of
chirality. The application of GC - functions in general
(\ref{m24}) and (\ref{m28}) forms lets to interpret these
operators in terms of the Pauli isospace vectors constructed on the
base of $\kappa_{x}, \ \kappa_{y}, \ \kappa_{z}$ connected
with x, y, z - axes consequently. Pauli transformations are
interpreted in the case as rotations depended on $\theta$ and $\chi$
angles. The former ones introduce Majorana terms into the Lagrangian
and into the generalized charge (\ref{m22}), while the latter, that
incorporate $\chi$ angle, describe the chiral transformations and
introduce the violation of ' - parity. The correlation of the general
form of Pauli Lagrangian with modern Majorana models \cite{11, 12},
 \cite{29} shows that it fits a special system including left - and
right - handed particles of the same flavor with the complex Dirac mass
($M_{D} = M \cos{\theta} e^{i\chi}$) and left and right Majorana
masses. In Pauli model under investigation the latter ones are equal in
magnitude but are of opposite signs ($M_{R} = - M_{L} =
 M \sin{\theta} e^{i\chi}$). The value $M^{2} = |M_{L(R)}|^{2} +
|M_{D}|^{2}$ is invariant of the group of general Pauli (chiral - Pauli)
transformations.

\par Thus, the use of Pauli transformations as a theoretical base
for a description of a neutral particle leads to a special Majorana
model. In the frame of the model the general Majorana scheme is
reduced to a special case that fits to one with an arbitrary Dirac
mass and left and right Majorana masses being opposite in signs but
equal in measure. Besides, in the frame of the model, in contrast
with general Majorana phenomenological schemes, there is a possibility
to generalize the concept of lepton charge and to link it with the form
of the Lagrangian mass term. In this way the eigenfunctions of
the generalized lepton charge operator and that of the mass structure
term can be used for description of arbitrary Dirac and Majorana
states of the Pauli model. In the case, the projection conditions
for eigenfunctions of basic operators take the form of Majorana
relations of general shape, determined by the inversion classes
of investigated particles. The Dirac case is an exception from
the general rule. In view of independence of its Lagrangian
from the inversion classes of the particles there are two
alternative representations of the solutions in the case. They
are described by either eigenfunctions of the lepton charge operator
or by those that define the structure of the Lagrangian mass term.
This alternative meets well known "Dirac" or "Majorana"
representations of the solutions. The model is described in the
isospace of Pauli transformations so that the basic operators,
determining the special representations, have properties of
spatial vectors.

\hspace{2mm}
\par

{\bf $\S 3.$ Two-flavor Pauli model of Majorana neutrinos}

\par Let us use the results obtained for the construction of a model
describing two neutrinos of different flavor. For definiteness we
shall assume that they are neutrino of electron (e) and muon ($\mu$)
flavors. Suppose that for the former case the left-handed state is
stands for the particle ($\nu_{e L}$) and its right-handed state does
for the antiparticle ($\nu^{C}_{e R}$). For the latter case the opposite
is true: the right-handed state is the particle ($\nu_{\mu R}$) and
its left-handed state is the antiparticle ($\nu^{C}_{\mu L}$). One can
coordinate the model with the above developed scheme of $\S 2$ by
using the substitutions:
\begin{equation}
\begin{array}{c}
\psi_{L}(x) \to \nu_{e L}(x), \ \
\psi_{R}(x) \to \nu_{\mu R}(x), \\
\psi^{C}_{R}(x) \to \nu^{C}_{e R}(x), \ \
\psi^{C}_{L}(x) \to \nu^{C}_{\mu L}(x).
\end{array}
\label{m31}
\end{equation}
\noindent Note that such a model is similar to Zel'dovich - Konopinsky -
Mahmoud scheme (ZKM - scheme) \cite{31, 32}, that was proposed by them
previously for the description of charged $e^{-}, \mu^{+}$ leptons. Such
schemes for neutrinos were described, for example in \cite{10}.

\par In accordance with general results of Pauli \cite{18} such a model
for massless neutrinos is invariant under chiral - Pauli transformations:
\begin{equation}
\begin{array}{c}
\nu^{'}_{e L}(x) = e^{i\chi/2} e^{i\varphi/2}

[\cos{(\theta/2)} \nu_{e L}(x) +
\sin{(\theta/2)} e^{-i\phi} \nu^{C}_{\mu L}(x) ], \\

\nu^{'}_{\mu R}(x) = e^{-i\chi/2} e^{i\varphi/2}

[\cos{(\theta/2)} \nu_{\mu R}(x) -
\sin{(\theta/2)} e^{-i\phi} \nu^{C}_{e R}(x) ], \\

\nu^{'C}_{e R}(x) = e^{-i\chi/2} e^{-i\varphi/2}
[\cos{(\theta/2)} \nu^{C}_{e R}(x) +
\sin{(\theta/2)} e^{+i\phi} \nu_{\mu R}(x) ], \\

\nu^{'C}_{\mu L}(x) = e^{i\chi/2} e^{-i\varphi/2}
[\cos{(\theta/2)} \nu^{C}_{\mu L}(x) -
\sin{(\theta/2)} e^{+i\phi} \nu_{e L}(x) ].
\end{array}
\label{m32}
\end{equation}
\noindent These transformations are canonical ones. In this
instance they mix electron and muon neutrino states of particle and
antiparticle types of general chirality with conserving commutation
relations of the massless fields.

\par The move to the two-flavor scheme by using (\ref{m31})
substitutions leads simultaneously to the changes of the standard
definitions of discrete C -, P -, T - transformations \cite{22} to
the following forms:
\begin{equation}
\begin{array}{c}
[\nu_{e L}(x)]^{C} = \nu_{e R}^{C}(x), \ \
[\nu_{\mu R}(x)]^{C} = \nu_{\mu L}^{C}(x), \\

[\nu_{e L}(Px)]^{(P)} = \eta_{P}\gamma_{4}\nu_{\mu R}(x), \ \
[\nu_{\mu R}(Px)]^{(P)} = \eta_{P}\gamma_{4}\nu_{e L}(x), \\
(P\vec{x} = - \vec{x}, \ Px_{4} = x_{4}), \\

[\nu_{e L}(Tx)]^{(T)} = - \eta_{T} \gamma_{2} \bar\nu_{e L}^{T}(x),  \ \
[\nu_{\mu R}(Tx)]^{(T)} = \eta_{T} \gamma_{2} \bar\nu_{\mu R}^{T}(x), \\
(T\vec{x} = \vec{x}, \ Tx_{4} = - x_{4}).
\end{array}
\label{m33}
\end{equation}
\noindent In the frame of these new definitions all results of above
investigated model due to properties of inversion classes of the
particles hold true in the two flavor neutrino scheme.

\par The most general Lagrangian of the two-flavor model for massive
neutrino, obtained from (\ref{m22}) by (\ref{m31}) substitutions,
has the following form:
\begin{equation}
\begin{array}{c}
L(x) = L_{0}(x) -  \frac{M}{2} \{\cos{\theta}
(\overline{\nu}_{\mu R}(x) e^{+i\chi} \nu_{e L}(x) +
\overline{\nu}_{\mu L}^{C}(x) e^{-i\chi} \nu^{C}_{e R}(x) + \\
\overline{\nu}_{e L}(x) e^{-i\chi} \nu_{\mu R}(x)
+ \overline{\nu}_{e R}^{C}(x) e^{+i\chi} \nu^{C}_{\mu L}(x)) + \\

\sin{\theta}(
\overline{\nu}_{\mu R}(x)e^{+i(\chi - \phi)} \nu_{\mu L}^{C}(x) +
\overline{\nu}^{C}_{\mu L}(x) e^{-i(\chi - \phi)} \nu_{\mu R}(x) - \\

\overline{\nu}^{C}_{e R}(x)e^{+i(\chi + \phi)} \nu_{e L}(x) -
\overline{\nu}_{e L}(x)e^{-i(\chi + \phi)} \nu_{e R}^{C}(x) ) \} \\

= L_{0}(x) - \frac{M}{2} \{\cos{\theta}
(\overline{\nu}_{\mu R}(x) e^{+i\chi} \nu_{e L}(x) +
\overline{\nu}_{\mu L}^{GC}(x) e^{+i\chi} \nu^{GC}_{e R}(x) + \\
\overline{\nu}_{e L}(x) e^{-i\chi} \nu_{\mu R}(x) +
\overline{\nu}_{e R}^{GC}(x) e^{-i\chi} \nu^{GC}_{\mu L}(x))

+ \sin{\theta}(\overline{\nu}_{\mu R}(x) \nu_{\mu L}^{GC}(x) + \\
\overline{\nu}^{GC}_{\mu L}(x) \nu_{\mu R}(x) -
\overline{\nu}^{GC}_{e R}(x) \nu_{e L}(x)  -
\overline{\nu}_{e L}(x) \nu_{e R}^{GC}(x) ) \}, \\

(\nu_{e L}(x))^{GC} = \nu_{e R}^{GC}(x) =
e^{-i(\chi + \phi)}\nu_{e R}^{C}(x), \\

(\nu_{\mu R}(x))^{GC} = \nu_{\mu L}^{GC}(x) =
e^{+i(\chi - \phi)}\nu_{\mu L}^{C}(x).
\end{array}
\label{m34}
\end{equation}
\noindent
\par The peculiarity of the Pauli neutrino scheme is in existence of a
conserved generalized lepton charge that is as follows for the
Lagrangian (\ref{m34}):
\begin{equation}
\begin{array}{c}
Q^{P}  =  \frac{1}{2}\int d^{3}x [\cos{\theta}
(\nu^{+}_{e L}(x)\nu_{e L}(x) -
\nu^{GC +}_{\mu L}(x)\nu^{GC}_{\mu L}(x) +
\nu^{+}_{\mu R}(x) \nu_{\mu R}(x)  - \\
\nu^{GC +}_{e R}(x)\nu^{GC}_{e R}(x)) +

\sin{\theta} (\nu^{+}_{e L}(x) e^{-i\chi} \nu^{GC}_{\mu L}(x) +

\nu^{GC +}_{\mu L}(x) e^{+i\chi} \nu_{e L}(x) - \\
\nu^{+}_{\mu R}(x) e^{+i\chi} \nu^{GC}_{e R}(x)  -
\nu^{GC +}_{e R}(x) e^{-i\chi} \nu_{\mu R}(x))]
\end{array}
\label{m35}
\end{equation}
\noindent Another important feature of the Pauli scheme is the
in\-va\-ri\-ance of the Lag\-ran\-gi\-an (\ref{m34}) un\-der
trans\-for\-ma\-tion:
\begin{equation}
\begin{array}{c}
\nu_{e L}(x) \to (\nu_{e L}(x))^{GC}, \ \
\nu^{GC}_{e R}(x) \to \nu_{e L}(x), \\

\nu_{\mu R}(x) \to (\nu_{\mu R}(x))^{GC}, \ \
\nu^{GC}_{\mu L}(x) \to \nu_{\mu R}(x),
\end{array}
\label{m36}
\end{equation}
\noindent that reverses the sign of the generalized lepton charge
(\ref{m35}). This transformation is equivalent to the operation of
the generalized charge (GC - ) conjugation (\ref{m23}) of the previous
model. It transforms the electron (left) and muon (right) neutrinos
into their own antiparticles with taking into account that the phase
factors $\eta_{C}$ of GC - operation can be different for electron
and muon neutrino flavor. The latter assumption is considered by
introduction of chiral $\chi \ne 0$ angle. As it was already underlined
earlier in the context of introduction of GC - operation (see discussion
 after (\ref{m23})) this difference is connected with CP - symmetry
violation.

\par The Lagrangian (\ref{m34}) describes a particular case of a
two-flavor Majorana neutrino system of ZKM - type, including a
left-handed electron neutrino with $M_{L}(\nu_{e})$ Majorana mass and
a right- handed muon neutrino  with $M_{R}(\nu_{\mu})$ Majorana mass.
As a consequence of Pauli-like scheme they are linked by the
expression:  $M_{R}(\nu_{\mu}) = - M_{L}(\nu_{e}) = M\sin{\theta}$.
These neutrinos can be mixed in between to be described by the Lagrangian
mass terms of "quasi-Dirac" type depended on $M_{D}(\nu_{\mu} \nu_{e}) =
M\cos{\theta} e^{i\chi}$ parameter.  The effective masses of the based
electron and muon neutrino states, which arise in the process of
diagonalization of the Lagrangian (\ref{m34}), are equal in magnitude and
opposite in signs:
\begin{equation}
\begin{array}{c}
M(\nu_{1}, \nu_{2}) =
\mp \sqrt{\frac{1}{4}(M_{R}(\nu_{\mu}) - M_{L}(\nu_{e}))^2
+ |M_{D}(\nu_{\mu} \nu_{e})|^{2}} = \mp M,
\end{array}
\label{m37}
\end{equation}
\noindent Thus, the length of the neutrino oscillation between the states
is equal to infinity. The mixing angle $\theta_{mix}$ introduced in a
standard way \cite{11} - \cite{12}, \cite{28} is given by the ratio
of "quasi - Dirac" to Majorana masses:
\begin{equation}
\begin{array}{c}
\tan{(2\theta_{mix})} = \cot{\theta} =
\frac{2 |M_{D}(\nu_{\mu} \nu_{e})|}
{M_{R}(\nu_{\mu}) - M_{L}(\nu_{e})}
= \frac{|M_{D}(\nu_{\mu} \nu_{e})|}{M_{R}(\nu_{\mu})}.
\end{array}
\label{m38}
\end{equation}

\par It is evident that among Lagrangians (\ref{m34}) it is
always possible to choose a basic one so that the others can be obtained
by proper general Pauli (chiral and Pauli) transformations. In the model
of one particle it was the Dirac Lagrangian, on whose basis the two
main representations of eigenfunctions were introduced in $\S 2$.
Those were the charge and Majorana (or mass) representations \cite{15, 16}.
Their basic operators were respectively the lepton charge operator
$\hat{\kappa}_{z}\gamma_{5}$ and $\hat{\kappa}_{z}$ operator, describing
the Lagrangian mass term. In two-flavor neutrino model these basic
operators are connected with different Lagrangians. Actually, in the
standard phe\-no\-me\-no\-lo\-gi\-cal schem\-es the re\-pre\-sen\-ta\-tion
with ab\-sence of mix\-ture of basic Majorana neutrino fields is
usually chosen as the basic one. In the neutrino model under investigation
such a representation is built on the basic $\hat{\kappa}_{z}$ operator
that is consistent with the choice of parameter $\theta = \eta \pi/2 \
(\theta_{mix} = 0) $ in (\ref{m34}). In this case the Lagrangian and
the generalized lepton charge obtain the following forms:
\begin{equation}
\begin{array}{c}
L(x) = L_{0}(x) + \frac{M}{2} \overline{\nu}_{0}(x)
\hat{\kappa}_{z} \nu_{0}(x) =  L_{0}(x) + \\ \frac{M}{2}
\{\overline{\nu}_{0 e}(x) \nu_{0 e}(x)
- \overline{\nu}_{0 \mu}(x) \nu_{0 \mu}(x) \}
=  L_{0}(x) +  \frac{M \eta}{2}
\{\overline{\nu}_{0 e L}(x)\nu_{0 e R}^{GC}(x) + \\
\overline{\nu}^{GC}_{0 e R}(x) \nu_{0 e L}(x)  -

\overline{\nu}_{0 \mu R}(x) \nu_{0 \mu L}^{GC}(x)  -
\overline{\nu}^{GC}_{0 \mu L}(x) \nu_{0 \mu R}(x) \}, \\

Q^{P} =  \frac{1}{2} \int d^{3}x
\nu_{0}^{+}(x) (\cos{\chi}\hat{\kappa}_{x} +
\sin{\chi}\hat{\kappa}_{y})
\gamma_{5} \nu_{0}(x) =  \\

 \frac{\eta}{2} \int d^{3}x
[\nu_{0 e L}^{+}(x) e^{-i\chi} \nu^{GC}_{0 \mu L}(x) +
\nu^{GC +}_{0 \mu L}(x) e^{+i\chi} \nu_{0 e L}(x) - \\
\nu^{+}_{0 \mu R}(x) e^{+i\chi} \nu^{GC}_{0 e R}(x)   -
\nu^{GC +}_{0 e R}(x) e^{-i\chi} \nu_{0 \mu R}(x) ],
\end{array}
\label{m39}
\end{equation}
\noindent Here the two - component basic neutrino function
$\nu_{0}(x)$ is as follows:
\begin{equation}
\begin{array}{c}
\nu_{0}(x) = {\nu_{0 e}(x) \choose  \nu_{0 \mu}(x)}, \\

\nu_{0 e}(x) =
\nu_{0 e L}(x) + \eta \nu_{0 e R}^{GC}(x) =

\nu_{0 e L}(x)+ \eta e^{-i(\chi + \phi)} \nu_{0 e R}^{C}(x),  \\

\nu_{0 \mu}(x) =
\nu_{0 \mu R}(x) + \eta \nu_{0 \mu L}^{GC}(x) =
\nu_{0 \mu R}(x) + \eta e^{+i(\chi - \phi)} \nu_{0 \mu L}^{C}(x).
\end{array}
\label{m40}
\end{equation}
\noindent It is similarly to (\ref{m12}) modified with a transition
to the GC - conjugation operation, and includes a pair of basic Majorana
solutions which are the electron $\nu_{0 e}(x)$ and muon $\nu_{0 \mu}(x)$
neutrinos with no mixing. The solutions have the general charge
parity $\eta = \pm 1$ due to the GC - conjugation and differ from each
other by the eigenvalues of $\hat{\kappa}_{z}$ operator. The latter is
an operator of neutrino flavor which simultaneously defines the structure
of the Lagrangian (\ref{m39}) mass term:
\begin{equation}
\begin{array}{c}
\hat{\kappa}_{z} \nu_{0 \ \kappa}(x) =
\kappa \nu_{0 \ \kappa}(x), \\

\nu_{0 \ +1}(x)
= {\nu_{0 e}(x) \choose  0}, \ \ (\kappa = +1), \ \

\nu_{0 \ -1}(x)
= {0 \choose  \nu_{0 \mu}(x)}, \ \ (\kappa = -1).
\end{array}
\label{m41}
\end{equation}
\noindent The corresponding equations in two-component and conventional
forms are as follows:
\begin{equation}
\begin{array}{c}
\gamma_{\mu} \partial_{\mu} \nu_{0}(x) -
M \hat{\kappa}_{z} \nu_{0}(x) = 0, \\

\gamma_{\mu} \partial_{\mu} \nu_{0 e L}(x) -
M \eta \nu^{GC}_{0 e R}(x) = 0, \ \

\gamma_{\mu} \partial_{\mu} \nu_{0 \mu R}(x) +
M \eta \nu^{GC}_{0 \mu L}(x) = 0, \\

\gamma_{\mu} \partial_{\mu} \nu^{GC}_{0 e R}(x) -
M \eta \nu_{0 e L}(x) = 0, \ \

\gamma_{\mu} \partial_{\mu} \nu^{GC}_{0 \mu L}(x) +
M \eta \nu_{0 \mu R}(x) = 0,
\end{array}
\label{m42}
\end{equation}
\noindent It is necassary to underline that the lower pair of the equations
is GC - conjugated to the previous one. Note that the equivalent equations
which link the left - handed particle (electron neutrino) and right -
handed antiparticle (electron antineutrino) components were obtained and
investigated in $\chi = \phi = 0, \ \ \eta = +1$ limit by Case \cite{30}.
In the case he considered the right - handed particle and left - handed
antiparticle (muon) components ignored, so that he actually kept on
investigating the $\kappa = +1$ solution with additionally imposed
conditions $\nu_{0 \mu R}(x) = \nu^{GC}_{0 \mu L}(x) = 0$.

\par For the description of the basic representation of the charge
type with an operator of the lepton charge as basic one it is useful to
conserve the Dirac form of the Lagrangian. It is due to full mixing
of electron and muon neutrino components of the generalized
wave function ($\theta = 0, \ \theta_{mix} = \pi/4$ in (\ref{m34}))
and it leads to the following forms of Lagrangian and lepton charge:
\begin{equation}
\begin{array}{c}
L(x) = L_{0}(x) - \frac{M}{2}
\overline{\nu}_{D}(x) (\cos{\chi}\hat{\kappa}_{x}
+ \sin{\chi} \hat{\kappa}_{y}) \nu_{D}(x) = \\
L_{0}(x) - \frac{M}{2}
(\overline{\nu}_{D \mu R}(x) e^{+i\chi}\nu_{D e L}(x)  +
\overline{\nu}^{GC}_{D \mu L}(x) e^{+i\chi} \nu^{GC}_{D e R}(x)
+ \\ \overline{\nu}_{D e L}(x) e^{-i\chi}\nu_{D \mu R}(x)  +
\overline{\nu}^{GC}_{D e R}(x) e^{-i\chi} \nu^{GC}_{D \mu L}(x)), \\

Q^{L} = \frac{1}{2} \int d^{3}x
\nu^{+}_{D}(x) \hat{\kappa}_{z} \gamma_{5}\nu_{D}(x)  =
\frac{1}{2} \int d^{3}x

(\nu^{+}_{D e L}(x) \nu_{D e L}(x)   + \\
\nu^{+}_{D \mu R}(x) \nu_{D \mu R}(x)  -

\nu^{GC +}_{D \mu L}(x) \nu^{GC}_{D \mu L}(x)  -
\nu^{GC +}_{D e R}(x) \nu^{GC}_{D e R}(x)), \\

\nu_{D}(x) = {\nu_{D e}(x) \choose \nu_{D \mu}(x) } =
{\nu_{D e L}(x) + \eta \nu^{GC}_{D e R}(x)
\choose \nu_{D \mu R}(x) + \eta \nu^{GC}_{D \mu L}(x) }.
\end{array}
\label{m43}
\end{equation}
\noindent The special case $\chi = 0$ is due to the real mass value
and is similar to (\ref{m18}). One can construct it from the latter
by using (\ref{m31}) substitutions in a way that it could be called as
the "quasi - Dirac" case. The eigenfunctions of the solutions with
fixed lepton charge $q_{z} = \pm 1$ are as follows:
\begin{equation}
\begin{array}{c}
\hat{\kappa}_{z} \gamma_{5} \nu_{D, q_{z}}(x)  =
q_{z} \nu_{D, q_{z}}(x), \\

\nu_{D, q_{z} = +1}(x) = {\nu_{D e L}(x) \choose \nu_{D \mu R}(x) }, \ \

\nu_{D, q_{z} = -1}(x) =

{\eta \nu^{GC}_{D e R}(x) \choose \eta \nu_{D \mu L}^{GC}(x) },
\end{array}
\label{m44}
\end{equation}
\noindent Note that in accordance with ZKM - scheme the lepton
charge of the left-handed electron and right-handed neutrinos is
equal to +1 while that for the right-handed electron and left-handed muon
antineutrinos is equal to -1.

\par It is interesting to note that one can obtain an alternative
description of quasi - Dirac case in flavor variables when using
Pauli transformation to $\theta' = - \pi/2$ angle of the wave
functions and operators of (\ref{m39}) representation.
The reason is connected with the conservation of the
flavor operator as the basic one although it takes a new form
$-(\cos{\chi}\hat{\kappa}_{x} + \sin{\chi}\hat{\kappa}_{y})$
(see (\ref{m43})) due to a nondiagonalized representation.
The generalized function of quasi - Dirac representation $\nu_{D}(x)$
takes the form:
\begin{equation}
\begin{array}{c}
\nu_{D}(x) =
e^{- i(\cos{\chi}\hat{\kappa}_{y} - \sin{\chi}\hat{\kappa}_{x})
\theta'/2} \nu_{0}(x) =
e^{+ i(\cos{\chi}\hat{\kappa}_{y} - \sin{\chi}\hat{\kappa}_{x})\pi/4}
\nu_{0}(x)  = \\

\frac{1}{\sqrt{2}} {1 \ \ + e^{-i\chi} \choose  - e^{+i\chi} \ \ 1}
{\nu_{0 e}(x) \choose  \nu_{0 \mu}(x)} =

\frac{1}{\sqrt{2}} {\nu_{0 e L}(x) + e^{-i\chi} \nu_{0 \mu R}(x)
+ \eta \nu_{0 e R}^{GC}(x) + \eta e^{-i\chi} \nu_{0 \mu L}^{GC}(x)
\choose
\nu_{0 \mu R}(x) - e^{i\chi} \nu_{0 e L}(x)
+ \eta \nu_{0 \mu L}^{GC}(x) - \eta e^{i\chi} \nu_{0 e R}^{GC}(x)},  \\

\nu_{D e L}(x) = \frac{1}{\sqrt{2}} (\nu_{0 e L}(x)
+ \eta e^{-i\chi} \nu_{0 \mu L}^{GC}(x)), \\
\nu_{D \mu R}(x) = \frac{1}{\sqrt{2}} (\nu_{0 \mu R}(x)
- \eta e^{i\chi} \nu_{0 e R}^{GC}(x)) \\

\nu^{GC}_{D e R}(x) = \frac{1}{\sqrt{2}} (\nu_{0 e R}^{GC}(x) +
\eta e^{-i\chi} \nu_{0 \mu R}(x)) \\
\nu^{GC}_{D \mu L}(x) = \frac{1}{\sqrt{2}}(\nu_{0 \mu L}^{GC}(x)
- \eta e^{i\chi} \nu_{0 e L}(x)).
\end{array}
\label{m45}
\end{equation}
\noindent The generalized flavor $\kappa = \pm 1$ serves here as a
quan\-tum cha\-rac\-te\-ris\-tic of the al\-ter\-na\-ti\-ve
re\-pre\-sen\-ta\-tion. Its eigenfunctions can be obtained from
(\ref{m40}) by using the same transformation:
\begin{equation}
\begin{array}{c}
- (\cos{\chi}\hat{\kappa}_{x} + \sin{\chi}\hat{\kappa}_{y})
\nu_{D, \kappa}(x)  = \kappa \nu_{D, \kappa}(x), \\

\nu_{D, \kappa = +1}(x) =
{\nu_{D e L}(x) + \eta \nu_{D e R}^{GC}(x)
\choose \nu_{D \mu R}(x) + \eta \nu^{GC}_{D \mu L}(x)}_{\kappa = +1} =

\frac{1}{\sqrt{2}}{\nu_{0 e L}(x) + \eta \nu_{0 e R}^{GC}(x)
\choose
- e^{i\chi}(\nu_{0 e L}(x) + \eta \nu_{0 e R}^{GC}(x))}, \\

\nu_{D, \kappa = -1}(x) =
{\nu_{D e L}(x) + \eta \nu_{D e R}^{GC}(x)
\choose \nu_{D \mu R}(x) + \eta \nu^{GC}_{D \mu L}(x)}_{\kappa = -1} =

\frac{1}{\sqrt{2}}{e^{-i\chi}(\nu_{0 \mu R}(x)

+ \eta \nu^{GC}_{0 \mu L}(x)) \choose
\nu_{0 \mu R}(x) + \eta \nu^{GC}_{0 \mu L}(x)},
\end{array}
\label{m46}
\end{equation}
\noindent The supplementary conditions of projection type, when choosing
definite values of the generalized flavor $\kappa$, can be presented
in separate or combined forms:
\begin{equation}
\begin{array}{c}
\nu_{D \mu L}^{C}(x) = - \eta e^{i\phi} \nu_{D e L}(x), \ \
\nu_{D e R}^{C}(x) = - \eta e^{i\phi} \nu_{D \mu R}(x) \ \

(\kappa = +1) , \\

\nu_{D \mu L}^{C}(x) = \eta e^{i\phi} \nu_{D e L}(x), \ \
\nu_{D e R}^{C}(x) = \eta e^{i\phi} \nu_{D \mu R}(x)  \ \
(\kappa = -1), \\

\nu_{D \mu}(x) =  - \kappa e^{i\chi} \nu_{D e}(x), \ \
(\kappa = \pm 1).
\end{array}
\label{m47}
\end{equation}
\noindent The latter expression means that the electron and muon
Majorana components of the Dirac neutrino of ZKM - scheme in the
states of fixed generalized flavor coincide in magnitude with accuracy
to the phase. The solutions (\ref{m44}) and (\ref{m46}) are examples
of an alternative description of the quasi - Dirac case in the two-flavor
neutrino Pauli model in terms of either lepton charge of ZKM - type or
generalized flavor.

\par Let us now move to the general case of the two-flavor model.
Starting from Lagrangian for the basic case of the flavor representation
$\theta = \eta \pi/2$ (\ref{m39}) and using Pauli transformation
one can construct the general expression of the two-component wave
function:
\begin{equation}
\begin{array}{c}
\nu(x)
= {\nu_{e}(x) \choose  \nu_{\mu}(x)} =

{\nu_{e L}(x) + \eta \nu^{GC}_{e R}(x)
\choose
\nu_{\mu R}(x) + \eta \nu^{GC}_{\mu L}(x)} =

e^{-i(\cos{\chi} \hat{\kappa}_{y} -

\sin{\chi} \hat{\kappa}_{x}) \theta'/2} \nu_{0}(x) = \\

{\cos{(\theta'/2)}\nu_{0 e}(x) - \sin{(\theta'/2)}
e^{-i\chi} \nu_{0 \mu}(x)
\choose
\cos{(\theta'/2)}\nu_{0 \mu}(x) + \sin{(\theta'/2)}
e^{+i\chi} \nu_{0 e}(x)}, \ \

(\theta' = \eta \theta - \pi/2).
\end{array}
\label{m48}
\end{equation}
\noindent The $\theta'$ parameter of the Pauli transformation can be
connected with the mixing angle introduced in a standard
phenomenological description of two-flavor Majorana models
\cite{10} - \cite{12}. It sets a degree of deviation of the investigated
Lagrangian from the basic one, in which electron and muon components of
the generalized function are not mixed. In this case the Lagrangian and
the conserved generalized charge have the following forms:
\begin{equation}
\begin{array}{c}
L(x) =  L_{0}(x) + \frac{M}{2}
\overline{\nu}(x) [\cos{\theta'} \hat{\kappa}_{z} +
\sin{\theta'}(\cos{\chi}\hat{\kappa}_{x} + \sin{\chi} \hat{\kappa}_{y})]
\nu(x),  \\

Q^{P} = \frac{1}{2} \int d^{3}x
\nu^{+}(x) [\cos{\theta'}(\cos{\chi}\hat{\kappa}_{x} +
\sin{\chi} \hat{\kappa}_{y}) - \sin{\theta'} \hat{\kappa}_{z}]
\gamma_{5}\nu(x),
\end{array}
\label{m49}
\end{equation}
\noindent The corresponding system of equations for the separated
components is as follows:
\begin{equation}
\begin{array}{c}
\gamma_{\mu} \partial_{\mu} \nu_{e L}(x) -
M \sin{\theta'} e^{-i\chi} \nu_{\mu R}(x)
- M \eta \cos{\theta'} \nu^{GC}_{e R}(x) = 0, \\

\gamma_{\mu} \partial_{\mu} \nu_{\mu R}(x) -
M \sin{\theta'} e^{+i\chi} \nu_{e L}(x)
+ M \eta \cos{\theta'} \nu^{GC}_{\mu L}(x) = 0, \\

\gamma_{\mu} \partial_{\mu} \nu^{GC}_{\mu L}(x) -
M \sin{\theta'} e^{+i\chi}\psi^{GC}_{e R}(x)
+ M \eta \cos{\theta'} \nu_{\mu R}(x) = 0, \\

\gamma_{\mu} \partial_{\mu} \nu^{GC}_{e R}(x) -
M \sin{\theta'} e^{-i\chi}\nu^{GC}_{\mu L}(x)
- M \eta \cos{\theta'} \nu_{e L}(x) = 0.
\end{array}
\label{m50}
\end{equation}
\noindent In the two-component form it is reduced to the equation:
\begin{equation}
\begin{array}{c}
\gamma_{\mu} \partial_{\mu} \nu(x) -
M \sin{\theta'} (\cos{\chi}\hat{\kappa}_{x}  +
\sin{\chi}\hat{\kappa}_{y}) \nu(x)
- M \cos{\theta'} \hat{\kappa}_{z} \nu(x) = 0.
\end{array}
\label{m51}
\end{equation}
\noindent It is useful to note that the conserved current
$J^{P}_{\mu}(x)$ has the following general form in the case:
\begin{equation}
\begin{array}{c}
J^{P}_{\mu}(x) =  \frac{i}{2}
\overline{\nu}(x) \gamma_{\mu} \gamma_{5}
[\cos{\theta'}(\cos{\chi}\hat{\kappa}_{x} + \sin{\chi} \hat{\kappa}_{y})
- \sin{\theta'} \hat{\kappa}_{z}] \nu(x)  = \\

\frac{i}{2}
[\cos{\theta'} (\overline{\nu}_{e L}(x) \gamma_{\mu} \nu^{C}_{\mu L}(x) +
\overline{\nu}^{C}_{\mu L}(x) \gamma_{\mu} \nu_{e L}(x)  -
\overline{\nu}_{\mu R}(x) \gamma_{\mu} \nu^{C}_{e R}(x) - \\
\overline{\nu}^{C}_{e R}(x) \gamma_{\mu} \nu_{\mu R}(x))
- \sin{\theta'}
(\overline{\nu}_{e L}(x) \gamma_{\mu} \nu_{e L}(x) +
\overline{\nu}_{\mu R}(x) \gamma_{\mu} \nu_{\mu R}(x) - \\

\overline{\nu}^{C}_{\mu L}(x) \gamma_{\mu} \nu^{C}_{\mu L}(x)  -
\overline{\nu}^{C}_{e R}(x) \gamma_{\mu} \nu^{C}_{e R}(x))].
\end{array}
\label{m52}
\end{equation}
\noindent It consists of two terms: the neutral current (proportional
to $\sin{\theta'}$) that does not change flavor number, and the current
which changes neutrino flavor from electron to muon type and vice versa
(proportional to $\cos{\theta'}$). The latter includes $\nu_{\mu} \to
\nu_{e} \ (\Delta \kappa = + 2)$ transitions as well as transitions of
$\nu_{e} \to \nu_{\mu} \ (\Delta \kappa = - 2)$ type.

\par By using expression (\ref{m48}) one can construct the general
form of the eigenfunctions with the generalized flavor operator
describing the structure of the Lagrangian mass term as a basic one.  The
eigenfunctions with fixed quantum number $\kappa = \pm 1$ take the form:
\begin{equation}
\begin{array}{c}
[\cos{\theta'} \hat{\kappa}_{z} + \sin{\theta'}
(\cos{\chi}\hat{\kappa}_{x} + \sin{\chi} \hat{\kappa}_{y})]
\nu_{\kappa}(x) = \kappa \nu_{\kappa}(x), \\

\nu_{\kappa = +1}(x)
= {\nu_{e}(x) \choose  \nu_{\mu}(x)}_{\kappa = +1} =

{\cos{(\theta'/2)} \nu_{0 e}(x) \choose
\sin{(\theta'/2)} e^{i\chi} \nu_{0 e}(x)} \ \ (\kappa = +1), \\
\nu_{\kappa = -1}(x)
= {\nu_{e}(x) \choose  \nu_{\mu}(x)}_{\kappa = -1} =

{-\sin{(\theta'/2)} e^{-i\chi} \nu_{0 \mu}(x) \choose
\cos{(\theta'/2)} \nu_{0 \mu}(x)} \ \ (\kappa = -1).
\end{array}
\label{m53}
\end{equation}
\noindent Their chiral components satisfy the relations:
\begin{equation}
\begin{array}{c}
\nu^{C}_{\mu L}(x)
= \kappa \eta (\tan{(\theta'/2)})^{\kappa} e^{i\phi} \nu_{e L}(x),  \\

\nu^{C}_{e R}(x)
= \kappa \eta (\cot{(\theta'/2)})^{\kappa} e^{i\phi} \nu_{\mu R}(x),
\ \ (\kappa =  \pm 1),
\end{array}
\label{m54}
\end{equation}
\noindent with the following combined form:
\begin{equation}
\begin{array}{c}
\nu_{\mu, \kappa}(x)
= \kappa (\tan{(\theta'/2)})^{\kappa} e^{i\chi} \nu_{e, \kappa}(x),
\ \ (\kappa =  \pm 1).
\end{array}
\label{m55}
\end{equation}
\noindent This relation generalizes expression (\ref{m47}) that
was found above for the quasi - Dirac case. It follows that in states of
fixed generalized flavor the contributions of Majorana electron and muon
neutrino components are bound together. In the case of the basic Lagrangian
(\ref{m39}) the contribution of muon constituent in the state of generalized
electron flavor $(\kappa = 1)$ is equal to zero. And in general case it is
proportional to the tangent of half an angle $\theta'$ which is responsible
for flavor mixing. The similar value estimates the contribution of Majorana
electronic constituent in the neutrino state of the generalized muon
flavor $(\kappa = -1)$. In "quasi - Dirac" states these contributions
coinside in magnitude.

\par In the two-flavor ZKM - model investigated these relations are
analogous to Majorana relations of projection types of $\S 2$ model.
The analysis of their behavior under space inversion shows that in
premise of universality of phases of space inversion transformations
they are realized for particles of inversion A-B - classes only
$(\eta_{P}^{2} = -1)$. However if electron and muon neutrinos belong
to different inversion classes it is allowed that electron neutrinos are
referred to C - class $(\eta_{P (e)} = +1)$ and muon neutrinos to D - class
$(\eta_{P (\mu)} = -1)$ or vice versa $(\eta_{P (e)} = -1, \ \eta_{P (\mu)}
= +1)$.

\par In the alternative description with generalized lepton charge
be\-ing us\-ed as a quan\-tum cha\-rac\-te\-ris\-tic its operator is
chosen as basic. In this case the Lagrangian (\ref{m43}), in which Pauli
rotation through angle $\theta' = -\pi/2$ relatively to (\ref{m39}) has
been already executed, becomes the basic one, so that for coordination
with the previous case it is necessary to make an additional rotation
through the angle $\eta \theta$. Once the rotation is performed
the following Lagrangian and the generalized lepton charge of (\ref{m24})
type will arise:
\begin{equation}
\begin{array}{c}
L(x) = L_{0}(x) -
\frac{M}{2} \overline{\nu'}(x) [\cos{\theta}(\cos{\chi}\hat{\kappa}_{x}+
\sin{\chi} \hat{\kappa}_{y}) - \eta\sin{\theta}\hat{\kappa}_{z}] \nu'(x), \\

Q^{L} = \frac{1}{2} \int d^{3}x \nu'^{+}(x)
[\cos{\theta}\hat{\kappa}_{z} + \eta \sin{\theta}
(\cos{\chi}\hat{\kappa}_{x} + \sin{\chi} \hat{\kappa}_{y})]
\gamma_{5}\nu'(x).
\end{array}
\label{m56}
\end{equation}
\noindent Note that expressions (\ref{m43}) and (\ref{m56}) are bound
by Pauli transformation:
\begin{equation}
\begin{array}{c}
\nu'(x)  = e^{-i\eta (\cos{\chi} \hat{\kappa}_{y} -
\sin{\chi}\hat{\kappa}_{x}) \frac{\theta}{2}} \nu_{D}(x) =

{\cos{(\theta/2)} \ \ -\eta \sin{(\theta/2)} e^{-i\chi}
\choose  \eta \sin{(\theta/2)} e^{+i\chi} \ \ \cos{(\theta/2)}}
{\nu_{D e}(x) \choose  \nu_{D \mu}(x)}  = \\

{\cos{(\theta/2)} \nu_{D e L}(x) -
\eta \sin{(\theta/2)} e^{-i\chi}\nu_{D \mu R}(x)
+ \eta \cos{(\theta/2)} \nu_{D e R}^{GC}(x)
- \sin{(\theta/2)} e^{-i\chi} \nu_{D \mu L}^{GC}(x)
\choose
\cos{(\theta/2)} \nu_{D \mu R}(x) +
\eta \sin{(\theta/2)} e^{+i\chi} \nu_{D e L}(x)
+ \eta \cos{(\theta/2)} \nu_{D \mu L}^{GC}(x)
+ \sin{(\theta/2)} e^{+i\chi} \nu_{D e R}^{GC}(x))}.

\end{array}
\label{m57}
\end{equation}
\noindent The eigenfunctions of particles with fixed generalized
lepton charge $q = \pm 1$ are obtained from (\ref{m44}) by using
the same transformations. They take the form:
\begin{equation}
\begin{array}{c}
[\cos{\theta} \hat{\kappa}_{z} + \eta \sin{\theta}
(\cos{\chi}\hat{\kappa}_{x} + \sin{\chi} \hat{\kappa}_{y})]\gamma_{5}
\nu'_{q}(x) = q \nu'_{q}(x), \\

\nu'_{q = +1}(x)
= {\nu'_{e}(x) \choose  \nu'_{\mu}(x)}_{q = +1} = \\

{\cos{(\theta/2)}\nu_{D e L}(x)
- \eta \sin{(\theta/2)} e^{-i\chi}\nu_{D \mu R}(x)
\choose
\cos{(\theta/2)}\nu_{D \mu R}(x) +
\eta \sin{(\theta/2)} e^{i\chi} \nu_{D e L}(x)} \ \ (q = +1), \\

\nu'_{q = -1}(x)
= {\nu'_{e}(x) \choose  \nu'_{\mu}(x)}_{q = -1} = \\

{\eta \cos{(\theta/2)}\nu^{GC}_{D e R}(x)
- \sin{(\theta/2)} e^{-i\chi} \nu^{GC}_{D \mu L}(x)
\choose
\eta \cos{(\theta/2)} \nu^{GC}_{D \mu L}(x)
+  \sin{(\theta/2)}e^{i\chi}\nu^{GC}_{D e R}(x)} \ \ (q = -1),
\ \ (\eta\theta = \theta' + \pi/2).
\end{array}
\label{m58}
\end{equation}
\noindent Their components satisfy the following projection Majorana
conditions similar to (\ref{m27}):
\begin{equation}
\begin{array}{c}
\nu'^{C}_{\mu L}(x)
= q (\tan{(\frac{\theta}{2})})^{q}
e^{i\phi} \nu'_{e L}(x), \\

\nu'^{C}_{e R}(x)
= - q (\tan{(\frac{\theta}{2})})^{q}
e^{i\phi} \nu'_{\mu R}(x) \ \ (q= \pm 1).
\end{array}
\label{m59}
\end{equation}
\noindent They connect electron neutrino components with charge - conjugated
muon ones and muon neutrino components with charge - conjugated electron ones.
It is evident that in the states of the fixed generalized lepton charge
q and chirality, the contribution of the states with the opposite lepton
charge for small angles $\theta \ (\theta' \sim -\pi/2)$ is small. However
it grows proportionally to the tangent of half an angle $\theta$ value
with receding from quasi - Dirac case and reaches equality with basic
contribution for $\theta = \eta \pi/2, \ (\theta' = 0) $, when Majorana
solutions of fixed flavor due to Lagrangian (\ref{m39}) exist. In this case
the angle $\theta$ plays a role of a parameter, describing the mixture of
particle and antiparticle states of quasi - Dirac types for solutions
of fixed generalized lepton charge.

\par From the analysis of behavior of expressions (\ref{m59}) under
space reflection it follows that in the case of universality of phase
factors $\eta_{P}$ they are fulfilled for particles of the inversion
C-D - classes $(\eta_{P}^{2} = +1)$ only. However if neutrinos are of
different inversion classes they can be satisfied also for particles of
inversion A-B - classes in the following cases: $\eta_{P (e)} = i$ (A -
class), $\eta_{P (\mu)} = -i$  (B - class) or $\eta_{P (e)} = -i$ (B -
class), $\eta_{P (\mu)} = +i$ (A - class). As a whole the expressions
(\ref{m54}) and (\ref{m59}) show that in the general case of the two-flavor
model the severe proportion of contributions of electron and muon
neutrino components exist in eigenfunctions of basic operators.
They play here the same role as projection Majorana relations do in
the above one-particle model.

\par Thus we have demonstrated that in the frame of the Pauli concept it
is possible to construct a two-flavor neutrino model, which incorporates
two sorts of massive neutrinos of different flavor and has quantum
characteristics of Zel'dovich - Konopinsky - Mahmoud type. The general
Lagrangian of such a model includes the mass terms connected with Majorana
masses of the neutrinos as well as the terms responsible for their mixing
(quasi - Dirac ones). In this case Majorana masses of different neutrinos
are equal in magnitude but differ in signs, so that the length of neutrino
oscillation between them tends to infinity. Such Pauli model is a special
case of general models of ZKM - type but has a set of important properties
which are absent in the general schemes. The existence of special quantum
numbers, universally describing either Majorana or quasi - Dirac states of
the model, is the main feature. In accordance with the choice of basic
operators two following possible alternative representations can be
realized. They are either flavor basic operator, which simultaneously
defines the mass structure of the specific Lagrangian, or the generalized
lepton charge, describing all the states of the system by using these
characteristics. With all this going on the choice of the basic operator is
dependent on the inversion class of a particle under investigation. There
is an exclusion from the general law which is the quasi - Dirac case, with
the alternative use of both representations being possible. The chiral -
Pauli transformation, transfering the basic form of definite representation
into the general one, includes the angle parameter, that sets the proper
rotation in the isospace of Pauli transformations as well as degree of
mixture of basic solutions of the basic form in a solution with the fixed
quantum number of the generalized type.

  However, the application of this scheme to the analysis of the experimental
data on neutrino oscillation needs taking into proper account an
inequality of Majorana electron and muon neutrino masses. Below we
shall show that a ne\-ces\-sa\-ry ex\-ten\-sion of the sche\-me presented
can be gi\-ven with con\-ser\-va\-tion of the peculiarities of the Pauli
scheme, provided that the proper general suppositions of the system
neutrino Lagrangian are adopted.

\par
\par
{\bf $\S 4.$ Neutrino oscillations in two-flavor Pauli model}

\par Let us consider the most general Lagrangian of phenomenological
Majorana models presented for example in \cite{11} and set off the term
of (\ref{m34}) type, which can be described by the two-flavor Pauli scheme.
Transforming it to notaions of the model by (\ref{m31}) substitutions
we can get the model Lagrangian:
\begin{equation}
\begin{array}{c}
L_{md}(x) =  L_{0}(x) -  \frac{1}{2} \{
\overline{\nu}_{\mu R}(x) m_{D} \nu_{e L}(x) + \\
\overline{\nu}_{e R}^{C}(x) m_{D} \nu^{C}_{\mu L}(x)
+ \overline{\nu}_{e L}(x) m^{*}_{D} \nu_{\mu R}(x)
+ \overline{\nu}_{\mu L}^{C}(x) m^{*}_{D} \nu^{C}_{e R}(x) \\

+ \overline{\nu}_{\mu R}(x)(m_{1} - i m_{2}) \nu_{\mu L}^{C}(x)  +
\overline{\nu}^{C}_{\mu L}(x)(m^{*}_{1} + i m^{*}_{2}) \nu_{\mu R}(x)  \\
+ \overline{\nu}^{C}_{e R}(x)(m_{1} + i m_{2}) \nu_{e L}(x)  +
\overline{\nu}_{e L}(x)(m^{*}_{1} - i m^{*}_{2}) \nu_{e R}^{C}(x) ) \}.
\end{array}
\label{m60}
\end{equation}
\noindent It depends on three phenomenological mass parameters
$m_{1}, \ m_{2}, \ m_{D}$, which are arbitrary complex values
usually connected with Majorana - $m_{1}, \ m_{2}$, and Dirac
(quasi-Dirac) - $m_{D}$ masses. As it was demonstrated above the
Pauli scheme is a special case of a general Majorana one
in which Majorana mass terms of L and R types are equal in value
and of different signs. For comparision with Pauli (\ref{m34}) form
let us connect the mass parameters of (\ref{m60}) with ones of the
two-flavor neutrino model of $\S 3$ by using the following relations:
\begin{equation}
\begin{array}{c}
m_{1} -i m_{2} = M_{R}(\nu_{\mu}) e^{i(\chi - \phi)}
= M'_{R}(\nu_{\mu}) e^{-i \phi}, \
m_{D} = M_{D}(\nu_{\mu} \nu_{e}) e^{+ i \chi}, \\

m_{1} +i m_{2} = M_{L}(\nu_{e}) e^{i(\chi + \phi)} =
M'_{L}(\nu_{e}) e^{+ i\phi}, \
\ m^{*}_{D} = M^{*}_{D}(\nu_{\mu} \nu_{e}) e^{- i \chi}, \\

Re M_{D}(\nu_{\mu} \nu_{e}) = M \cos{\theta} , \ \
\frac{1}{2}(M_{R}(\nu_{\mu}) - M_{L}(\nu_{e})) = M \sin{\theta},
\end{array}
\label{m61}
\end{equation}
\noindent If in the further course one uses in these expressions
the neutrino GC - functions (\ref{m34}) then the model
Lagrangian (\ref{m60}) can be rewritten in the form:
\begin{equation}
\begin{array}{c}
L_{md}(x) =  L_{0}(x) -
\frac{M}{2} [ \cos{\theta}
(\overline{\nu}_{\mu R}(x) e^{i\chi} \nu_{e L}(x) + \\
\overline{\nu}_{\mu L}^{GC}(x) e^{i\chi} \nu^{GC}_{e R}(x)
+ \overline{\nu}_{e L}(x) e^{-i\chi}\nu_{\mu R}(x)
+ \overline{\nu}_{e R}^{GC}(x) e^{-i\chi} \nu^{GC}_{\mu L}(x)) + \\

\sin{\theta} (\overline{\nu}_{\mu R}(x) \nu_{\mu L}^{GC}(x)  +
\overline{\nu}^{GC}_{\mu L}(x) \nu_{\mu R}(x)
- \overline{\nu}_{e L}(x) \nu_{e R}^{GC}(x) - \\
\overline{\nu}^{GC}_{e R}(x) \nu_{e L}(x)) ]

- \frac{1}{2} \{
\frac{M_{R}(\nu_{\mu}) + M_{L}(\nu_{e})}{2}
(\overline{\nu}_{\mu R}(x) \nu_{\mu L}^{GC}(x)  + \\
\overline{\nu}^{GC}_{\mu L}(x) \nu_{\mu R}(x)
+ \overline{\nu}_{e L}(x) \nu_{e R}^{GC}(x)
+ \overline{\nu}^{GC}_{e R}(x) \nu_{e L}(x)) + \\

\frac{M_{D}(\nu_{\mu} \nu_{e}) - M^{*}_{D}(\nu_{\mu} \nu_{e}) }{2}
(\overline{\nu}_{\mu R}(x) e^{i\chi} \nu_{e L}(x)
+ \overline{\nu}_{e R}^{GC}(x) e^{-i\chi} \nu^{GC}_{\mu L}(x) - \\
\overline{\nu}_{\mu L}^{GC}(x) e^{i\chi} \nu^{GC}_{e R}(x)

- \overline{\nu}_{e L}(x) e^{-i\chi}\nu_{\mu R}(x)) \},
\end{array}
\label{m62}
\end{equation}
\noindent The second term of the expression is in coincidence with
the mass term of the Lagrangian (\ref{m34}) and two latter ones comprise
the additive mass term which takes into account the inequality of
Majorana masses of electron and muon neutrinos and possible complexity
of the quasi - Dirac $M_{D}(\nu_{\mu} \nu_{e})$ parameter. Note, that the
term, depending on half a sum of Majorana masses of the electron and
muon neutrinos, is symmetric under operation of the generalized charge
GC - conjugation as the Pauli Lagrangian (\ref{m34}) does. It allows
in some way to simplify the form (\ref{m62}) by implying the
similar symmetry condition to the whole Lagrangian of the model as a
general requirement. For $Im M_{D}(\nu_{\mu} \nu_{e}) = 0$ the
Lagrangian does not incorporate the latter term but however it is in
consistency with a rather general Majorana model for two-flavor neutrinos
with unequal Majorana masses as before. Since GC - conjugation takes into
account different phase factors $\eta_{C}$ for neutrinos of different
flavors, the introduction of a similar general symmetry condition under
that operation can be considered as natural and physically non
contradictory for the general Majorana scheme.

\par The two-flavor netrino oscillations have a finite oscillation
length in the modified general model. Indeed, let us introduce
a $\nu_{md}(x)$ neutrino function that is an analogue of $\nu(x)$
(\ref{m48}) in the Lagrangian of the model, and conduct a transformation,
leading to reduction of the Pauli mass term of the Lagrangian to the
form (\ref{m39}). To realize it one has to transform the shortened
Lagrangian (\ref{m62}) to its general form (\ref{m49}) and then perform
the correspondent Pauli transormation with using rotation inversed to
(\ref{m48}):
\begin{equation}
\begin{array}{c}

L_{md}(x) =  L_{0}(x)- \frac{M_{0}}{2}\overline{\nu}_{md}(x) \nu_{md}(x)
 + \frac{M}{2} \overline{\nu}_{md}(x)

\hat{\kappa} \nu_{md}(x), \\

M_{0} = \frac{M_{R}(\nu_{\mu}) + M_{L}(\nu_{e})}{2} \eta, \ \

\hat{\kappa} = \cos{\theta'} \hat{\kappa}_{z} +
\sin{\theta'}(\cos{\chi}\hat{\kappa}_{x} +
\sin{\chi}\hat{\kappa}_{y}),
\end{array}
\label{m63}
\end{equation}
\noindent It reduces the Lagrangian to the following diagonal form:
\begin{equation}
\begin{array}{c}
L_{md}(x) =  L_{0}(x)- \frac{M_{0}}{2} \overline{\nu}_{md0}(x) \nu_{md0}(x)
+ \frac{M}{2}\overline{\nu}_{md0}(x) \hat{\kappa}_{z} \nu_{md0}(x), \\

\nu_{md0}(x) = {(\nu_{md0})_{e}(x) \choose (\nu_{md0})_{\mu}(x)}, \ \

\nu_{md0}(x) = e^{i(\cos{\chi}\hat{\kappa}_{y} -
\sin{\chi}\hat{\kappa}_{x})\theta'/2} \nu_{md}(x), \\

\gamma_{\mu} \partial_{\mu} \nu_{md0}(x) + M_{0} \nu_{md0}(x)
- M \hat{\kappa}_{z} \nu_{md0}(x) = 0, \
\theta' = \eta \theta - \pi/2.
\end{array}
\label{m64}
\end{equation}
\noindent In accordance with the standard procedure of Majorana
phenomenology (see, e.g. \cite{11, 12}) physical neutrinos entered into
$\nu_{md}(x)$ are superpositions of the basic neutrino states of definite
masses $(\nu_{md0})_{e (\mu)}(x)$. The latter ones obey equations
(\ref{m64}) and are described by the universal flavor number $\kappa$ of
the two-flavor model of the previous section. In passing from Lagrangian
(\ref{m63}) to the basic form the effective Majorana masses of electron and
muon neutrino take the values:
\begin{equation}
\begin{array}{c}
M(\nu_{e})
= M_{0} - M = \\ \eta \frac{M_{R}(\nu_{\mu}) + M_{L}(\nu_{e})}{2} -
\sqrt{\frac{1}{4}(M_{R}(\nu_{\mu}) - M_{L}(\nu_{e}))^2 +
(Re M_{D}(\nu_{\mu} \nu_{e}))^{2}}, \\

M(\nu_{\mu}) = M_{0} + M = \\
\eta \frac{M_{R}(\nu_{\mu}) + M_{L}(\nu_{e})}{2} +
\sqrt{\frac{1}{4}(M_{R}(\nu_{\mu}) - M_{L}(\nu_{e}))^2  +
(Re M_{D}(\nu_{\mu} \nu_{e}))^{2}}.
\end{array}
\label{m65}
\end{equation}
\noindent As a result, the known formulae arise for effective Majorana
masses of neutrino of two types, that define the length of their mutual
oscillations through the difference of their squared masses:
\begin{equation}
\begin{array}{c}
|M^{2}(\nu_{\mu}) - M^{2}(\nu_{e})| = 4 |M_{0}| M, \\
L_{osc} = 4\pi E/|M^{2}(\nu_{\mu}) - M^{2}(\nu_{e})|
= \pi E/ |M_{0}| M.
\end{array}
\label{m66}
\end{equation}
\noindent The new principal feature of the Pauli model, following
from the structure of Lagrangian (\ref{m64}) and mass formula
(\ref{m65}), is a fact that the effective masses of free Majorana
particles consist of two terms different in their nature. They have
properties of Pauli isoscalar and Pauli isovector. The former is universal
and depends on mass characteristics of neutrinos only, the latter is
described by the flavor quantum number $\kappa$ and depends on the mixing
angle $\theta'$. The universality of the former means that separate
components of wave function $\nu_{md}(x)$ can be determined by the same
quantum numbers of flavor type as components of $\nu(x)$ function of Pauli
model. Besides, such an interpretation shows that in the modified
two-flavor Pauli model the effective neutrino Majorana masses are formed
with participation of not one but two different Higgs scalar fields,
which also have properties of a scalar and a vector of the Pauli
isospace.

\par In parallel with the length of neutrino oscillations the other
important experimental parameter of physical neutrino is the mixing
angle. In phenomenological neutrino schemes it is included
on the base of the connection between wave functions of physical
left-handed neutrino states and eigenfunctions of states
$\nu_{1 L}(x), \ \nu_{2 L}(x)$ of fixed masses. It is described
by the standard relations:
\begin{equation}
\begin{array}{c}
\nu_{e L}(x) =
\nu_{1 L}(x) \cos{\theta_{mix}} + \nu_{2 L}(x)\sin{\theta_{mix}}, \\
\nu_{\mu L}(x)  = \nu_{2 L}(x) \cos{\theta_{mix}}
- \nu_{1 L}(x) \sin{\theta_{mix}}.
\end{array}
\label{m67}
\end{equation}
\noindent In the two-flavor Pauli model they appear as a consequence
of Pauli transformations (\ref{m64}) reducing Lagrangian (\ref{m63})
to a diagonal form with ei\-gen\-func\-tions con\-struc\-ted of the wave
func\-tions of fix\-ed mass\-es and fla\-vors $(\nu_{md0})_{e(\mu) \rho}(x),
\\ (\rho = L, R)$. The $\theta'$ angle that specifies these transitions, can
be coordinated with the mixing angle $\theta_{mix}$ defined from
(\ref{m67}) and can be deduced from the neutrino oscillation
experiments. It is obvious that the connections equivalent to
(\ref{m67}) are described in the model under investigation by
using a transformation of Pauli type, inversed to (\ref{m64}).
They are as follows:
\begin{equation}
\begin{array}{c}
\nu_{md}(x) =
e^{-i(\cos{\chi}\hat{\kappa}_{y} -
\sin{\chi}\hat{\kappa}_{x})\theta'/2} \nu_{md0}(x) = \\

{\cos{(\theta'/2)}(\nu_{md0})_{e}(x) - \sin{(\theta'/2)}
e^{-i\chi} (\nu_{md0})_{\mu}(x)
\choose
\cos{(\theta'/2)}(\nu_{md0})_{\mu}(x) + \sin{(\theta'/2)}
e^{+i\chi} (\nu_{md0})_{e}(x)},  \\

(\nu_{md})_{e L}(x) =
\cos{(\theta'/2)}(\nu_{md0})_{e L}(x) - \eta \sin{(\theta'/2)}
(\nu^{GC}_{md0})_{\mu L}(x), \\

(\nu^{GC}_{md})_{\mu L}(x)  =
\cos{(\theta'/2)}(\nu^{GC}_{md0})_{\mu L}(x)
+ \eta \sin{(\theta'/2)} (\nu_{md0})_{e L}(x), \\

(\nu_{md})_{\mu R}(x) =
\cos{(\theta'/2)}(\nu_{md0})_{\mu R}(x) + \eta \sin{(\theta'/2)}
(\nu^{GC}_{md0})_{e R}(x), \\

(\nu^{GC}_{md})_ {e R}(x) =
\cos{(\theta'/2)} (\nu^{GC}_{md0})_{e R}(x)
- \eta \sin{(\theta'/2)} (\nu_{md0})_{\mu R}(x), \\

\theta' = \eta \theta - \pi/2.
\end{array}
\label{m68}
\end{equation}
\noindent (The phase factor $e^{-i\phi} = \eta_{C}$ is included
in the definition of charge GC-conjugation operation). The relations
(\ref{m68}) are analogues of formulae (\ref{m67}) to present the
eigenfunctions of the physical neutrino through the states of a fixed
masses. The latter are basic flavor states of Pauli model and coincide
with (\ref{m67}) under the following conditions:
\begin{equation}
\begin{array}{c}
\nu_{e L}(x) = (\nu_{md})_{e L}(x), \
\nu_{\mu L}(x)  = (\nu^{GC}_{md})_{\mu L}(x), \

\nu_{1 L}(x) = (\nu_{md0})_{e L}(x), \\
\nu_{2 L}(x) = (\nu^{GC}_{md0})_{\mu L}(x), \

\tan{(2\theta_{mix})} = \eta \cot{\theta}, \\

2\theta_{mix} = - \theta' = \pi/2 - \eta \theta, \ \eta = +1 .
\end{array}
\label{m69}
\end{equation}
\noindent The first ones describe transition from the standard
scheme to Majorana one of Zel'dovich - Konopinsky - Mahmoud type,
the second ones state the connection between the experimental mixing
angle $\theta_{mix}$ and the $\theta$ angle of the Pauli scheme. The
former sets the direction of the vector of the generalized lepton
charge relatively to z - axis of the Pauli isospace for two neutrinos
of different flavors.

\par As an example of application of the model under investigation
we propose a qualitative interpretation of the experimental neutrino
oscillation results. The modern data on the mixing angles and the squared
neutrino mass differences \cite{33} are as follows:
\begin{equation}
\begin{array}{c}
(\theta_{mix})_{12} = (34 \pm 2.3)^{\circ}, \ \

(\theta_{mix})_{23} = (45 \pm 8.2)^{\circ}, \ \

(\theta_{mix})_{13} \le 13^{\circ}, \\

\Delta M_{12}^{2} = M_{1}^{2} - M_{2}^{2} \sim 8 \cdot 10^{-5} \ eV^{2}, \ \

\Delta M_{23}^{2} = M_{2}^{2} - M_{3}^{2} \sim 2.5 \cdot 10^{-3} \ eV^{2}.
\end{array}
\label{m70}
\end{equation}
\noindent It is evident that as a consequence of the small mixing angle
$(\theta_{mix})_{13}$ the physical neutrinos of electron and $\tau$
types can be approximately described in the frame of two-component
mixing. However the muon neutrino is a mixture of three components
and can not be described by such a simple model. For Pauli parameters
$\theta$, which set $\nu_{e} - \nu_{\mu}$ and $\nu_{\tau} - \nu_{\mu}$
mixtures, one can extract the following experimental values
by using (\ref{m69}):
\begin{equation}
\begin{array}{c}
(\theta_{exp})_{12} = (22 \pm 4.6)^{\circ}, \ \

(\theta_{exp})_{23} = (0 \pm 16)^{\circ}.
\end{array}
\label{m71}
\end{equation}
\noindent These values show that $\tau$ - neutrino is a "quasi-Dirac"
mixture of $\nu_{\tau L}$ and $\nu_{\mu R}$, which can be described
as the state of lepton charge of ZKM -type with $q = +1$.
In accordance with (\ref{m53}) the electron neutrino is described
as a mixture of Majorana states $\nu_{e, \kappa}$ and $\nu_{\mu, \kappa}$,
specifing by the generalized flavor number $\kappa = +1$. Using
(\ref{m55}) one can estimate that the fraction of the muon neutrino
Majorana component is about half of the electron neutrino fraction value.
The muon neutrino is a complex mixture of all three neutrino states
and is not described in our scheme. The three-flavor neutrino Majorana
model on the base of Pauli transformations will be presented
by the author in a special separate article.

\par

{\bf $\S 5.$ Conclusion}

\par Thus, the two models for the description of Majorana properties of
neutral free fermions have been constructed on the base of general
Pauli (chiral - Pauli) transformations. The first one is a model
for one particle with states of left and right chirality that
developes the simplest initial scheme proposed by Majorana \cite{1}
(see also \cite{30}). It extends the latter to the case when the mass
part of its Lagrangian includes the both Dirac and Majorana mass
terms. The second one describes a system of neutral particles of two
different flavors so that the particle state of the first one (conditionally
called as electron neutrino) has the left chirality and that of the other
(called as muon neutrino) has the right chirality, i.e. it goes in
consistence with the Zel'dovich - Konopinsky - Mahmoud scheme. The models
under investigation are special cases of general Majorana schemes presented
in literature.  The latter one can be modified and used for description of
the simplest, two flavor version of the neutrino oscillation in ZKM -
scheme.

\par As a consequence of connection with Pauli transformations
these Pauli schemes posess the following peculiarities:

\par 1. These models describe a special class of Majorana Lagrangians of
neutral particles connected by general Pauli (chiral - Pauli)
transformations, that lets one to get an arbitarary Lagrangian
by starting from the basic Lagrangian of definite
representation of the model. The chiral - Pauli transformations
consist of pure Pauli SU(2) - group and chiral group of U(1) - type.
For massless particles the general Pauli symmetry is exact,
however it is destroyed with introduction of mass terms in the
Lagrangian.

\par 2. In coordination with general suggestions of the Standard model
the mass terms of Pauli Lagrangians are interpreted as a result
of the spontaneous symmetry breakdown that pick out a special direction
in the space of Pauli transformations with nonzero mean vacuum
value of Higgs field. In assumption of universality of the breakdown
mechanism the mass terms of different Pauli Lagrangians are
connected with the same transformations as the correspondent
specified directions outligned above. The hypothesis of universality is
equivalent to the assumption that the Higgs field has the properties
of a vector in Pauli isospace. In this case the mean vacuum value
set fixed its modulus and specifies the effective mass of basic
particles, and the angle coordinates of the vector, defining
the relative values of Dirac and Majorana mass terms, are not changed
with the spontaneous breakdown. The spontaneously broken symmetry due to
chiral transformations is connected with the breakdown of CP - invariance.

\par The modified version of the two-flavor Pauli scheme introduces
an additional mass term of the Lagrangian with properties of
isoscalar of Pauli space, so that quantum characteristics of the
states due to isovector part are conserved in the modified scheme as well.
The Higgs field of the modified model consists of two components
related to Pauli isovector and isoscalar types of mass terms
of the Lagrangian.

\par 3. The conception of the lepton charge of neutral particles is
extended to the generalized lepton charge, that is coordinated with an
arbitrary direction of the Pauli isospace. For an arbitrary Lagrangian
the corresponding operator of the generalized lepton charge includes
parameters of Pauli rotations. On the base of this operator the
product of the generalized lepton charge and chirality that can
be used as an alternative basic operator for description of
neutral particles is constructed. The quantum numbers of these
operators allow to describe universally the states of Dirac
(quasi-Dirac) as well as those of pure Majorana or mixed Dirac -
Majorana types. The generalization of the lepton charge leads to a
modification of the operation of charge conjugation. In the
generalized charge GC - conjugation $\eta_{C}$ phase factors for
the left-handed and right-handed particles can be different;
that is connected with a violation of CP-symmetry.

\par 4. The basic representations of the models are given by their
Lagrangians and basic operators, the other representations are
connected with them by Pauli transformations. In the first
model it is Dirac Lagrangian with two basic representations:
the charge ("Dirac") one, where the solutions are described
by lepton charges, and the Majorana ("mass") one with two
Majorana solutions of fixed mass which differ in signs
of $\psi^{C}(x) = \pm \psi(x)$ condition. Their basic operators
are correspondingly either the operator of the lepton charge or
the operator of the structure of the Lagrangian mass term. The
latter is connected with the product of lepton charge and chirality. In
the basic "mass" ("flavor") representation of the second model the basic
neutrino Majorana states of electron and muon flavors are not mixed so that
the mean value of the lepton charge is equal to zero. The basic charge
representation of the second model is analogous of "Dirac" one
("quasi - Dirac") of the first model. It has the basic states of
fixed lepton charge of ZKM - type, which are the mixture of electron
and muon neutrino Majorana components with their contributions being
coincident in values.

\par 5. The interpretation of Pauli models depends on the
inversion classes of particles under investigation, which are
defined by the phase factors of the space inversion operation.
They are $\eta_{P} =  \pm i$ for particles of inversion A and
B - classes and $\eta_{P} = \pm 1$ for inversion C and D - classes
respectively. In modern models it is supposed that neutrinos
as well as the other physical particles belong to A-B - inversion
classes; however in Majorana neutrino schemes this hypothesis should
be controlled experimentally. In Pauli models the form of Majorana
conditions is connected with the inversion classes of particles.
For example, $\psi^{C}(x) = \pm \psi(x)$ condition is realized
for inversion A-B - classes only and its analogue for C-D -
classes is $\psi^{C}(x) = \pm \gamma_{5} \psi(x)$ condition.
In general case the form of Majorana conditions is generalized
but defined as above by inversion classes of particles under
investigation. This rigid connection between the form of Majorana
conditions and inversion class of particles can be destroyed if
one lets the phase factors $\eta_{P}$ be not universal, so that
different physical neutrino states could belong to different
inversion classes.

\par 6. The choice of the basic operator, its eigenfunctions and
projection conditions of fixed eigenvalue is connected
with the inversion class of particles under consideration. In
general case there are two alternative representations for
wave functions of the states: the mass (flavor) representation with
basic operator of the structure of Lagrangian mass term (generalized
flavor) and the charge representation, when the operator of the
generalized lepton charge is basic. These operators do not commutate
and are complementary. For universal definition of the phases of space
inversion operation, the first representation is realized for
particles of inversion A-B - classes and the second one is realized for
particles of inversion C-D - classes. If Lagrangian is not sensitive
to the inversion classes of particles, as for example in Dirac
("quasi - Dirac") case, the alternative description is possible
in every representation. The projection conditions for
eigenfunctions take either form of Majorana conditions (first model)
or the form of connection between electron and muon neutrino components of
particle and antiparticle type (second model). For particles of
inversion A-B - classes their forms generalize the known Majorana
conditions but for C-D - classes those include, in addition, the
$\gamma_{5}$ operator.

\par 7. The basic operator of the basic representation is a vector.
It is coordinated with z - axis of the Pauli isospace and its
eigenvalues are coordinated with projections of the vector to the
axis. For arbitrary directions the basic operators are coordinated
with vectors of the correspondent directions. For the representations
using the operator of the generalized lepton charge, the basic case
corresponds to Dirac (quasi - Dirac) description on the ground
of independent particle and antiparticle states and correlates with
z - axis. The general Pauli Lagrangian is characterized by $\theta$
angle between the vector of the generalized lepton charge and z - axis
of the Pauli isospace. The intermediate cases are connected with the
Dirac - Majorana mixture defined by the $\theta$ angle, and
the Majorana case corresponds to $\theta = \eta \pi/2$, when particle
and antiparticle contributions in eigenfunctions of the generalized
lepton charge are equal. For the representations using operator of the
generalized flavor ("mass") type, the mixture of Majorana states of
different flavor is absent in the basic representaion. The intermediate
cases are connected with incorporation of the mixture given by $\theta'$
angle. For pure Dirac (quasi - Dirac) case $\theta' = - \pi/2$, that
corresponds to maximal mixing of basic flavor ("mass") states.
In neutrino oscillation experiments the mixing angle is usually
measured starting from the pure Majorana case and it characterizes
the extend of mixture of basic, flavor Majorana states in the
state of physical neutrino.

\par 8. The mass part of Majorana Lagrangians contains terms of
two types: those that specify Majorana masses of the particles
$(M_{M1}, \ M_{M2})$ and those that specify their mixture $(M_{12})$.
The latter are interpreted in terms of mixture of Majorana particles
and associated with Dirac (quasi - Dirac) terms of the model. The Pauli
models are specified by the fact that Majorana mass terms of the
described particles (they are left and right ones - in the first model
or electron and muon neutrinos - in the second one respectively)
are equal in magnitude and opposite in signs $(M_{M2} = - M_{M1})$.
In the conventional method of diagonalization of Lagrangians with
transition to Majorana representation the former ones are set
to the terms which depend on effective masses of particles
$(M_{1}, \ M_{2})$. The latter ones tend to zero so that the states
of fixed effective masses are not mixed. In Pauli case the application of
diagonalization method is realized by chiral - Pauli transformations and
leads to the following effective masses of the particles:
\begin{equation}
\begin{array}{c}
M_{1, 2} = \pm \sqrt{\frac{1}{4}|M_{M1} - M_{M2}|^{2} +
|M_{12}|^{2}}, \ \ M_{M1} + M_{M2} = 0,
\end{array} \
\label{m72}
\end{equation}
\noindent In this case the lengths of oscillations between
electron and muon neutrino states of the second model occur to tend to
infinity:
\begin{equation}
\begin{array}{c}
L_{osc} = 4 \pi E/|M_{1}^2 -
M_{2}^2| = \infty.
\end{array}
\label{m73}
\end{equation}
\noindent In the modified neutrino Pauli scheme the additional universal
term is included into the Lagrangian of the system, so that the effective
masses of neutrinos differ in measure and the length of their mutual
oscillations becomes finite:
\begin{equation}
\begin{array}{c}
M_{1, 2} =
M_{0} \pm \sqrt{|M_{M2}|^{2} + |M_{12}|^{2}}, \ \

L_{osc} = \pi E/M_{0} \sqrt{|M_{M2}|^{2} + |M_{12}|^{2}}.
\end{array}
\label{m74}
\end{equation}
\noindent In this case the oscillation parameter $\Delta M^2$
depends on mass characteristics of the Pauli model only, which are
invariants of isovector and isoscalar terms of Pauli Lagrangian
and do not depend on the mixing angle.

\par The investigation of the models arises the following important
general question: "Do Majorana particles of inversion C-D - classes exist in
the nature?" The similar particles should have nonstandard properties under
an operation of the space inversion and special Majorana conditions
which include additional $\gamma_{5}$ operator, different from the
standard ones in existing Majorana schemes.

{\bf Acknowledgments}

\par The author is very grateful to E.P.Velikhov for his continued
interest and support, as well as to S.M.Bilenky, B.V.Da\-ni\-lin,
G.V.Do\-mo\-gat\-sky, D.I.Ka\-za\-kov, Yu.V.Lin\-de, V.A.Ru\-ba\-kov,
S.V.Se\-me\-nov and V.V.Khruschev for useful discussions. The work
is supported by Grant N 26 on fundamental researches of RRC
"Kurchatov Institute" of 2006 - 2007 years.

\end{document}